\documentclass{iopjournal-arxiv}

\usepackage{graphicx}%
\usepackage{multirow}%
\usepackage{amsmath,amssymb,amsfonts}%
\usepackage[title]{appendix}%
\usepackage{xcolor}%
\usepackage{orcidlink}
\usepackage{textcomp}%
\usepackage{manyfoot}%
\usepackage{booktabs}%
\usepackage{listings}%
\usepackage{bm}%
\usepackage{siunitx}%
\usepackage{float}

\begin{document}

\articletype{Paper}

\title{Correlation-based Modeling of Seismic Newtonian Noise in Half-Space and Full-Space Media}

\author{Mohamed Samy$^{1*}$\,\orcidlink{0009-0004-6424-9726}, Jan Harms$^{1,2}$\,\orcidlink{0000-0002-7332-9806} and Tomislav Andric$^{1,2}$\,\orcidlink{0000-0002-9277-9773}}

\affil{$^1$Gran Sasso Science Institute (GSSI), I-67100 L'Aquila, Italy}

\affil{$^2$INFN, Laboratori Nazionali del Gran Sasso, I-67100 Assergi, Italy}

\affil{$^*$Author to whom any correspondence should be addressed.}

\email{mohamedsamyabdelmotteleb.elzokm@gssi.it}



\keywords{Newtonian noise, gravity gradient noise, gravitoelastic tensor, Rayleigh waves, body waves, Einstein Telescope, noise cancellation, seismometer arrays}

\abstract{Seismic Newtonian noise, arising from fluctuating gravitational forces on detector test masses due to ambient seismic activity, represents a fundamental sensitivity limit for low-frequency gravitational-wave observatories such as the Einstein Telescope. Effective mitigation of Newtonian noise requires detailed knowledge of the statistical correlations between the Newtonian acceleration perturbation at the test mass and the seismic displacement field measured by surrounding sensor arrays. In this work, the gravitoelastic correlation tensors---the cross-correlations between the Newtonian acceleration perturbation and the seismic displacement field---are derived and numerically validated for Rayleigh waves and body waves in half-space and full-space media, considering test masses located above and below ground, with and without a spherical cavity. The analytical solutions provide exact and asymptotic benchmarks for validating a Cartesian numerical integration framework, which reproduces the corresponding gravitoelastic tensors across Rayleigh-wave and body-wave models, establishing a unified tool for Newtonian-noise modeling and sensor-array design in future gravitational-wave detectors.}



\section{Introduction}\label{sec:intro}

The ground-based gravitational-wave (GW) detectors Advanced Virgo and Advanced LIGO have opened a new era of GW and multi-messenger astronomy, beginning with the first direct detection GW150914 and followed by the multi-messenger event GW170817 \cite{GW150914,GW170817}. Together with KAGRA, they form a global network of ground-based interferometers \cite{KAGRA2021}. Next-generation observatories such as Cosmic Explorer and especially the Einstein Telescope (ET) aim to extend GW observations to the few-Hz regime \cite{ET2020,Evans2021}. In this low-frequency band, seismic Newtonian noise (SNN) produced by gravitational coupling between the environment and the suspended test masses of a detector is expected to be one of the main environmental limitations to the target sensitivity \cite{Harms2019,Beker2011}. 

Mitigation of Newtonian noise can be achieved by using data from a seismometer array to produce a coherent model of Newtonian noise, which is subtracted from the GW data in post-processing. This procedure is known as Newtonian-noise cancellation (NNC). It requires a precise model of the seismic field and the associated SNN to determine the optimal configuration of the seismometer array \cite{Badaracco2019,Badaracco_2020,Schillings_2025}. In absence of an observation of Newtonian noise, the main challenge is to construct a numerical model that implements all the relevant site characteristics like geology, topography, planned cavern and tunnel geometries, and which also reproduces the seismic observations including seismic spectra, Rayleigh-wave dispersion, and seismic correlations. 

There are different approaches to calculate NN based on such models. A time-domain integration of the seismic wavefield was implemented in SPECFEM3D \cite{Harms_2015}. Accurate estimates of NN resulting from an integration over the seismic displacement field require the simulation of large volumes so that the integral converges. Integrals over seismic fields in large models have larger numerical errors. For this reason, an alternative approach was conceived based on gravitoelastic correlations \cite{Driggers2012,Coughlin2016,Harms2019,Badaracco2019,AndricHarms2020}. 
Simple analytical models or numerical simulations provide gravitoelastic correlations on a relatively sparse grid, and integrals over the correlation fields have smaller numerical errors. The drawback is that correlations live in a higher-dimensional coordinate space (six dimensional instead of three dimensional), which may cause other challenges in later processing steps \cite{Badaracco_2020}. It still seems to be the numerically more robust approach, and it will serve as an important benchmark to verify other modeling techniques.

In this work, we develop a comprehensive numerical framework to compute the gravitoelastic correlation tensor across three-dimensional seismic sensor grids. Our formulation is adaptable, evaluating NN correlations for test masses located both above and below the ground, and it easily incorporates specific local geometries, such as a spherical cavity, for underground configurations. To systematically validate this numerical approach, we derive exact and leading-order asymptotic analytical expressions for Rayleigh waves and for body waves in both half-space and full-space homogeneous isotropic media. By benchmarking the numerical integration against these analytical solutions, we validate the framework across all tensor components and establish a unified methodology for modeling gravitoelastic correlations relevant to NN mitigation in future GW detectors.

The paper is organized as follows: Section~\ref{sec:ge_tensor_general} develops the general equations of the gravitoelastic correlation tensor in the integral form in terms of the displacement correlation tensor, sections~\ref{sec:half-space} and~\ref{sec:body_full} develop the closed forms of gravitoelastic tensors and displacement correlation tensors for half-space and full-space media, respectively, Section~\ref{sec:validation} presents the numerical validation, Section~\ref{sec:discussion} discusses the results, and Section~\ref{sec:conclusion} summarizes the main conclusions.

\section{General Gravitoelastic Correlation Formalism}\label{sec:ge_tensor_general}

In this section, we derive the general integral equations for the gravitoelastic correlation tensor in terms of the displacement correlation tensor.

\subsection{Newtonian acceleration}

Designing an optimal cancellation system requires a quantitative model of the statistical relationship between the Newtonian acceleration at the test masses and the seismic displacement measured by sensors placed at various locations in and around the detector site. This relationship is captured by the \emph{gravitoelastic correlation tensor}~\cite{Harms2019},
\begin{equation}
  c_{ij}(\mathbf{r},\omega)
    = \bigl\langle \delta a_i(\mathbf{r}_0,\omega)\,
                   \xi_j^*(\mathbf{r},\omega) \bigr\rangle,
  \label{eq:ge_tensor_def}
\end{equation}
which correlates the $i$-th component of the Newtonian acceleration perturbation $\delta a_i$ at the test-mass position $\mathbf{r}_0$ with the $j$-th component of the seismic displacement $\xi_j$ measured at an arbitrary sensor position $\mathbf{r}$. The brackets $\langle \cdot \rangle$ denote an ensemble average over realizations of the seismic field.

Throughout this work, frequency-domain fields use the temporal convention
\(\exp(-i\omega t)\), with \(\omega=2\pi f\). Correlation tensors are
understood as cross-spectral densities at a fixed positive frequency, and the
modal displacement spectra use a one-sided convention. A consistent change of
spectral convention rescales all correlations by the same overall factor and
does not affect the normalized validation errors.

The total perturbed gravitational potential for an arbitrary background density field $\rho(\mathbf{r}')$ is given by the sum of its bulk and surface contributions \cite{Harms2019}:
\begin{equation}
\delta\Phi(\mathbf{r}_0,\omega)
= G \int_V d^3r'\,
\frac{\partial'_\alpha(\rho(\mathbf{r}')\xi_\alpha(\mathbf{r}',\omega))}{|\mathbf{r}'-\mathbf{r}_0|}
 - G \int_S dS'\,
\frac{\rho(\mathbf{r}')n_\alpha(\mathbf{r}')\xi_\alpha(\mathbf{r}',\omega)}{|\mathbf{r}'-\mathbf{r}_0|}.
\label{eq:phi_surf}
\end{equation}

The Newtonian acceleration is $\delta a_i(\mathbf{r}_0,\omega) = -\partial_{0,i} (\delta\Phi)$. Evaluating the gradient with respect to the test mass coordinate $\mathbf{r}_0$ yields the general acceleration:
\begin{equation}
\delta a_i(\mathbf{r}_0,\omega)
= -G \int_V d^3r'\,
\frac{r'_i-r_{0,i}}{|\mathbf{r}'-\mathbf{r}_0|^3}
\partial'_\alpha(\rho(\mathbf{r}')\xi_\alpha)
+ G \int_S dS'\,
\frac{r'_i-r_{0,i}}{|\mathbf{r}'-\mathbf{r}_0|^3}
\rho(\mathbf{r}')n_\alpha\xi_\alpha,
\label{eq:acc_total_general}
\end{equation}
where all fields inside the integrals are evaluated at $(\mathbf r',\omega)$. The kernel is singular at $\mathbf r'=\mathbf r_0$; if $\mathbf r_0\in V$, the integral is understood in the Cauchy principal value sense. \\
When the background density is a homogeneous constant, $\rho(\mathbf{r}') = \rho_0$, the density gradient vanishes and $\rho_0$ factors out of the spatial derivatives and integrals:
\begin{equation}
\delta a_i(\mathbf{r}_0,\omega)
= -G\rho_0 \int_V d^3r'\,
\frac{r'_i-r_{0,i}}{|\mathbf{r}'-\mathbf{r}_0|^3}
(\partial'_\alpha\xi_\alpha)
+ G\rho_0 \int_S dS'\,
\frac{r'_i-r_{0,i}}{|\mathbf{r}'-\mathbf{r}_0|^3}
(n_\alpha\xi_\alpha).
\label{eq:acc_total_const}
\end{equation}

\subsection{Gravitoelastic correlation}
 
\subsubsection{Bulk--surface model}
Starting from Eq.~\eqref{eq:ge_tensor_def} and exchanging differentiation with ensemble averaging gives:
\begin{align}
c_{ij}(\mathbf r,\omega)
&=
-\,G
\int_V d^3r'\,
\frac{r'_i-r_{0,i}}{|\mathbf r'-\mathbf r_0|^3}
\,\partial'_\alpha (\rho(\mathbf{r}')C_{\alpha j}(\mathbf r',\mathbf r;\omega))
\nonumber\\
&\quad
+\,G
\int_S dS'\,
\frac{r'_i-r_{0,i}}{|\mathbf r'-\mathbf r_0|^3}
\,\rho(\mathbf{r}')n_\alpha(\mathbf r')\,C_{\alpha j}(\mathbf r',\mathbf r;\omega).
\label{eq:ge_bulk_surface_general}
\end{align}

Where $C_{\alpha j}$ is the two-point displacement correlation tensor defined as:
\begin{equation}
C_{\alpha j}(\mathbf r',\mathbf r;\omega)
=
\left\langle
\xi_\alpha(\mathbf r',\omega)\,
\xi_j^*(\mathbf r,\omega)
\right\rangle .
\label{eq:ge_displacement_def}
\end{equation}

For constant density, factoring out $\rho_0$, defining $D_j(\mathbf r',\mathbf r;\omega) = \partial'_\alpha C_{\alpha j}(\mathbf r',\mathbf r;\omega)$, and using $\mathbf n=\hat{\mathbf z}$ for a flat surface $z'=0$, we obtain:
\begin{align}
c_{ij}(\mathbf r,\omega)
&=
-\,G\rho_0
\int_V d^3r'\,
\frac{r'_i-r_{0,i}}{|\mathbf r'-\mathbf r_0|^3}
D_j(\mathbf r',\mathbf r;\omega)
\nonumber\\
&\quad
+\,G\rho_0
\int_{z'=0} d^2\rho'\,
\frac{r'_{S,i}-r_{0,i}}{|\mathbf r'_S-\mathbf r_0|^3}
C_{zj}\big((\boldsymbol\rho',0),\mathbf r;\omega\big).
\label{eq:bulk_surface_model_const}
\end{align}

\subsubsection{Kernel--contraction model}

A derivative-free form is obtained by transferring the spatial derivative from the displacement correlation tensor to the Newtonian kernel. Defining the vector kernel:
\begin{equation}
K_i(\mathbf r')
=
\frac{r'_i-r_{0,i}}{|\mathbf r'-\mathbf r_0|^3},
\end{equation}

and applying the product rule,
\begin{equation}
K_i\,\partial'_\alpha (\rho C_{\alpha j})
=
\partial'_\alpha(K_i \rho C_{\alpha j})
-
(\partial'_\alpha K_i)\,\rho C_{\alpha j},
\end{equation}
then applying the divergence theorem to the first term on the right-hand side yields a volume term plus boundary contributions. For the volume integral of the second term, we introduce the tensor kernel:
\begin{equation}
T_{i\alpha}(\mathbf r')
=
\frac{\delta_{i\alpha}}{|\mathbf r'-\mathbf r_0|^3}
-
3\,\frac{(r'_i-r_{0,i})(r'_\alpha-r_{0,\alpha})}{|\mathbf r'-\mathbf r_0|^5}.
\end{equation}

Because of the singularity at $\mathbf r'=\mathbf r_0$, the kernel derivative is understood distributionally as
\begin{equation}
\partial'_\alpha K_i
=
\operatorname{PV}\!\left[T_{i\alpha}\right]
+
\frac{4\pi}{3}\delta_{i\alpha}\delta^{(3)}(\mathbf r'-\mathbf r_0).
\end{equation}

The boundary $\partial V$ consists of the free surface $S$ and the boundary at infinity $S_\infty$. The contribution from $S$ exactly cancels the explicit surface term in Eq.~\eqref{eq:ge_bulk_surface_general}. Evaluating the delta-function contribution against the local background density $\rho(\mathbf{r}_0)$ yields:
\begin{equation}
c_{ij}(\mathbf r,\omega)
=
c_{ij}^{(\mathrm{PV})}(\mathbf{r},\omega)
+
\frac{4\pi}{3}G\rho(\mathbf{r}_0)\,C_{ij}(\mathbf r_0,\mathbf r;\omega)
-
G\int_{S_\infty} dS'\,
n_\alpha(\mathbf r')\,K_i(\mathbf r')\,
\rho(\mathbf{r}')C_{\alpha j}(\mathbf r',\mathbf r;\omega),
\end{equation}
where $c_{ij}^{(\mathrm{PV})}(\mathbf{r},\omega)$ is the principal-value (PV) form obtained by excluding an infinitesimal sphere of radius $a$ around the singularity and taking $a\to 0$:
\begin{equation}
  c_{ij}^{(\mathrm{PV})}(\mathbf{r},\omega)
    = \lim_{a\to 0}\,G
      \int_{|\mathbf{r}'-\mathbf{r}_0|>a}
      d^3r'\,T_{i\alpha}(\mathbf{r}')\,
      \rho(\mathbf{r}')C_{\alpha j}(\mathbf{r}',\mathbf{r};\omega).
  \label{eq:PV_def}
\end{equation}

If the contribution at infinity vanishes, the exact derivative-free representation becomes:
\begin{equation}
c_{ij}(\mathbf r,\omega)
=
c_{ij}^{(\mathrm{PV})}(\mathbf{r},\omega)
+
\frac{4\pi}{3}G\rho(\mathbf{r}_0)\,C_{ij}(\mathbf r_0,\mathbf r;\omega).
\label{eq:kernel_contraction_model_general}
\end{equation}

When the background density is constant, factoring $\rho_0$ out yields:
\begin{equation}
c_{ij}(\mathbf r,\omega)
=
c_{ij}^{(\mathrm{PV})}(\mathbf{r},\omega)
+
\frac{4\pi}{3}G\rho_0\,C_{ij}(\mathbf r_0,\mathbf r;\omega).
\label{eq:kernel_contraction_model_const}
\end{equation}
Where
\begin{equation}
  c_{ij}^{(\mathrm{PV})}(\mathbf{r},\omega)
    = G \rho_0 \lim_{a\to 0}\,
      \int_{|\mathbf{r}'-\mathbf{r}_0|>a}
      d^3r'\,T_{i\alpha}(\mathbf{r}')\,
      C_{\alpha j}(\mathbf{r}',\mathbf{r};\omega).
  \label{eq:PV_def_rho0}
\end{equation}

The local term contributes whenever $\mathbf r_0\in V$. This includes underground half-space and full-space configurations, but not an above-ground half-space test mass.

\subsection{Test mass with cavity}
For a test mass underground inside a spherical cavity, we need to exclude a spherical cavity around the test mass. The spherical cavity centered at the test mass is:
\begin{equation}
  B_a(\mathbf r_0)=\left\{\mathbf r'\in\mathbb R^3: \left| {\mathbf r'-\mathbf r_0}\right|<a\right\}.
\end{equation}
In everything below, \textbf{the cavity means source exclusion}. The observation point is not excluded.

\subsubsection{Bulk-surface model}
For the no-cavity problem, the total correlation can be written as $c_{ij}(\mathbf r,\omega)=c_{ij}^{\mathrm{bulk}}(\mathbf r,\omega)+c_{ij}^{\mathrm{surf}}(\mathbf r,\omega)$. With a cavity, the perforated domain produces an extra inner-boundary term:
\begin{equation}
\label{eq:route3-cavity-decomp}
  c_{ij}^{\mathrm{cav}}(\mathbf r,\omega)
  =c_{ij}^{\mathrm{bulk,cav}}(\mathbf r,\omega)
  +c_{ij}^{\mathrm{surf}}(\mathbf r,\omega)
  +c_{ij}^{\mathrm{wall}}(\mathbf r,\omega).
\end{equation}
The bulk term is integrated over $V\setminus B_a(\mathbf r_0)$, the surface term is over the free surface $S$, and the wall term is over the cavity sphere $\partial B_a$. For a small cavity, the wall term evaluates the flux against the local density at the cavity location, giving the leading-order approximation:
\begin{equation}
  \label{eq:wall-small-cavity}
  c_{ij}^{\mathrm{wall}}(\mathbf r,\omega)
  =G\rho(\mathbf{r}_0)\left[
  -\frac{4\pi}{3}C_{ij}(\mathbf r_0,\mathbf r;\omega)
  + O\!\left(\frac{a^2}{\lambda^2}C\right)\right].
\end{equation}
Hence the working bulk-surface cavity model approximation is
\begin{equation}
\label{eq:bs-cav}
  c_{ij}^{\mathrm{cav}}(\mathbf r,\omega)
  \approx
  c_{ij}^{\mathrm{bulk,cav}}(\mathbf r,\omega)
  +c_{ij}^{\mathrm{surf}}(\mathbf r,\omega)
  -\frac{4\pi}{3}G\rho(\mathbf{r}_0)\,C_{ij}(\mathbf r_0,\mathbf r;\omega).
\end{equation}
For a constant density, replace $\rho(\mathbf{r}_0)$ with $\rho_0$ in Eq.~\eqref{eq:bs-cav}.

\subsubsection{Kernel-contraction model}
By excluding a cavity around the test mass, the local singular term is completely removed and the source domain is restricted to $V\setminus B_a(\mathbf r_0)$. The counterpart of Eq.~\eqref{eq:kernel_contraction_model_general} is:
\begin{equation}
\label{eq:kc-cavity}
  c_{ij}^{\mathrm{cav}}(\mathbf r,\omega)
  =G\int_{V\setminus B_a(\mathbf r_0)}
  T_{i\alpha}(\mathbf r'-\mathbf r_0)
  \rho(\mathbf{r}')C_{\alpha j}(\mathbf r',\mathbf r;\omega)
  \,d^3r'.
\end{equation}
For a constant density, replace $\rho(\mathbf{r}')$ with $\rho_0$ in Eq.~\eqref{eq:kc-cavity}

\subsection{Projection along an arbitrary detector arm}
The gravitoelastic correlation tensor is initially computed in Cartesian form according to Eq.~\eqref{eq:ge_tensor_def}. For a detector arm oriented along a generic unit vector $\mathbf u = (u_x, u_y, u_z)$, the longitudinal acceleration and displacement components are $\delta a_{\mathrm{arm}} = u_i\,\delta a_i$ and $\xi_{\mathrm{arm}} = u_j\,\xi_j$, respectively. The projected gravitoelastic correlation along this arm is obtained via the double contraction of the tensor with the orientation vector:
\begin{equation}
c_{\mathrm{arm}}(\mathbf r,\omega)
=
u_i\,c_{ij}(\mathbf r,\omega)\,u_j.
\label{eq:c_arm_projection_updated}
\end{equation}
More generally, the cross-correlation between the acceleration projected along $\mathbf u$ and the displacement projected along a distinct vector $\mathbf v$ is given by
\begin{equation}
\left\langle
(\mathbf u\cdot\delta\mathbf a)\,(\mathbf v\cdot\boldsymbol\xi)^*
\right\rangle
=
u_i\,c_{ij}(\mathbf r,\omega)\,v_j.
\end{equation}
In the specific case where the detector arm lies in the horizontal plane at an azimuthal angle $\delta$ relative to the $x$-axis, the orientation vector is $\mathbf u = (\cos\delta, \sin\delta, 0)$. Eq.~\eqref{eq:c_arm_projection_updated} then expands to
\begin{equation}
c_{\mathrm{arm}}
=
c_{xx}\cos^2\delta + c_{yy}\sin^2\delta + (c_{xy}+c_{yx})\cos\delta\sin\delta.
\end{equation}

\section{Homogeneous isotropic half-space medium}
\label{sec:half-space}

\subsection{Displacement correlation tensors}

For two points at positions $\mathbf{r}'=(x',y',z')$ and $\mathbf{r}=(x,y,z)$, separated by an angle $\varphi$, we define the horizontal separation as:
\begin{equation}
\Delta\boldsymbol{\rho} = \boldsymbol{\rho}'-\boldsymbol{\rho} = (\Delta x,\Delta y), \qquad \Delta x=x'-x, \qquad \Delta y=y'-y,
\end{equation}
with magnitude and directional components:
\begin{equation}
\Delta\rho=|\Delta\boldsymbol{\rho}| = \sqrt{\Delta x^2+\Delta y^2}, \qquad \Delta x=\Delta\rho\cos\varphi, \qquad \Delta y=\Delta\rho\sin\varphi.
\end{equation}

For a plane-wave component propagating in the horizontal direction \(\theta\), we define
\begin{equation}
    \hat{\mathbf e}_h(\theta)
    =
    \cos\theta\,\hat{\mathbf e}_x
    +
    \sin\theta\,\hat{\mathbf e}_y .
\end{equation}

For horizontally isotropic fields, the displacement vector can be expressed in terms of a generalized baseline amplitude $s$ and generalized horizontal and vertical depth profiles, $D_h(z)$ and $D_z(z)$:
\begin{equation}
\boldsymbol{\xi}(\mathbf{r},\omega) = s e^{i\mathbf{k}_h\cdot\boldsymbol{\rho}} \left[ D_h(z)\hat{\mathbf{e}}_h + D_z(z)\hat{\mathbf{e}}_z \right].
\end{equation}

The two-point displacement correlation tensor is obtained by averaging over the azimuth $\theta$ of the horizontal wavevector $\mathbf{k}_h$:
\begin{equation}
C_{\alpha j}(\mathbf{r}',\mathbf{r}) = \frac{1}{2\pi} \int_0^{2\pi} \xi_\alpha(\mathbf{r}';\theta) \xi_j^*(\mathbf{r};\theta) \,d\theta.
\end{equation}

Using the standard angular Bessel identities in App.~\ref{app:bessel-identities}, this evaluates to the generalized two-point correlation tensor:
\begin{equation}
\mathbf{C}(\mathbf{r}',\mathbf{r}) = |s|^2 
\begin{pmatrix} 
\frac{1}{2}[J_0 - \cos 2\varphi J_2] D'_h (D_h)^* & -\frac{1}{2}\sin 2\varphi J_2 D'_h (D_h)^* & i\cos\varphi J_1 D'_h (D_z)^* \\[1.5em] 
-\frac{1}{2}\sin 2\varphi J_2 D'_h (D_h)^* & \frac{1}{2}[J_0 + \cos 2\varphi J_2] D'_h (D_h)^* & i\sin\varphi J_1 D'_h (D_z)^* \\[1.5em] 
i\cos\varphi J_1 D'_z (D_h)^* & i\sin\varphi J_1 D'_z (D_h)^* & J_0 D'_z (D_z)^* \end{pmatrix},
\label{eq:C_general}
\end{equation}
where $J_n \equiv J_n(\kappa \Delta\rho)$, and the depth profiles are evaluated at their respective coordinates such that $D'_h = D_h(z')$, $D'_z = D_z(z')$, $D_h = D_h(z)$, and $D_z = D_z(z)$. 

\subsubsection{Specific wave modes}

The two-point displacement-correlation tensor in
Eq.~\eqref{eq:C_general} is universal within the present horizontally
isotropic formulation. Its mode-specific forms are obtained by substituting
the corresponding displacement profiles \(D_h\) and \(D_z\), together with
the base displacement amplitude \(s\), as specified in
App.~\ref{app:master_equations}.

\subsection{Exact Closed-Form Gravitoelastic Tensor}
\label{ssec:closed_form_general}

For a horizontally isotropic wave field, consider an observation point
\begin{equation}
    \mathbf r=(x,y,z),
    \qquad
    \rho=\sqrt{x^2+y^2},
    \qquad
    x=\rho\cos\varphi,
    \qquad
    y=\rho\sin\varphi .
\end{equation}
The test mass is located on the vertical axis,
\begin{equation}
    \mathbf r_0=(0,0,z_0).
\end{equation}

The perturbed gravitational acceleration at the test-mass position is written as
\begin{equation}
    \delta\mathbf a(\mathbf r_0,\omega;\theta)
    =
    2\pi G\rho_0s
    \left[
        A_h(z_0)\hat{\mathbf e}_h(\theta)
        +
        A_z(z_0)\hat{\mathbf e}_z
    \right].
    \label{eq:general_acceleration_A}
\end{equation}

The gravitoelastic tensor is defined by the angular average
\begin{equation}
    c_{ij}(\mathbf r,\omega)
    =
    \frac{1}{2\pi}
    \int_0^{2\pi}
    \delta a_i(\mathbf r_0,\omega;\theta)
    \xi_j^*(\mathbf r,\omega;\theta)
    \,d\theta .
    \label{eq:ge_angular_avg}
\end{equation}

Using the Bessel identities in App.~\ref{app:bessel-identities}, the universal closed-form tensor becomes
\begin{equation}
\mathbf c(\mathbf r,\omega)
=
2\pi G\rho_0|s|^2
\begin{pmatrix}
\dfrac12
\left[J_0-\cos 2\varphi\,J_2\right]
A_h D_h^*
&
-\dfrac12
\sin 2\varphi\,J_2
A_h D_h^*
&
-i\cos\varphi\,J_1
A_h D_z^*
\\[1.2em]
-\dfrac12
\sin 2\varphi\,J_2
A_h D_h^*
&
\dfrac12
\left[J_0+\cos 2\varphi\,J_2\right]
A_h D_h^*
&
-i\sin\varphi\,J_1
A_h D_z^*
\\[1.2em]
-i\cos\varphi\,J_1
A_z D_h^*
&
-i\sin\varphi\,J_1
A_z D_h^*
&
J_0
A_z D_z^*
\end{pmatrix}.
\label{eq:c_general_AD}
\end{equation}
where $J_n \equiv J_n(\kappa\rho)$, and the profiles are evaluated at their respective coordinates such that $A_h=A_h(z_0)$, $A_z=A_z(z_0)$, $D_h^*=D_h^*(z)$, and $D_z^*=D_z^*(z)$.

\subsubsection{Specific wave modes}

The gravitoelastic tensor in Eq.~\eqref{eq:c_general_AD} is universal within
the present horizontally isotropic formulation. Its mode-specific forms are
obtained by substituting the corresponding acceleration amplitudes \(A_h\)
and \(A_z\), displacement profiles \(D_h\) and \(D_z\), and base displacement
amplitude \(s\), as specified in App.~\ref{app:master_equations}.

\subsection{Incoherent P--SV Modal Mixtures}\label{ssec:incoherent_mixture}

The simplest statistically meaningful extension from a pure-mode field to a
multimodal seismic field assumes that the \(P\)-wave and \(SV\)-wave
populations are statistically independent, or \emph{incoherent}. Let
\(\bm{\xi}^{P}\) and \(\bm{\xi}^{SV}\) denote modal fields normalized to the
same reference displacement power, and introduce nonnegative modal power
fractions \(w_P\) and \(w_{SV}\). A realization of the mixed field may then be
written as
\begin{equation}
\bm{\xi}^{\mathrm{mix}}
=\sqrt{w_P}\,\bm{\xi}^{P}
+\sqrt{w_{SV}}\,\bm{\xi}^{SV},
\qquad
w_P+w_{SV}=1.
\label{eq:mixed_field_realization}
\end{equation}
The corresponding acceleration field carries the same square-root amplitude
weights because the gravitational coupling is linear. The incoherence
assumption implies that all mixed cross-correlations vanish:
\begin{equation}
\bigl\langle \xi_\alpha^{P}(\mathbf r',\omega)\,\xi_j^{SV*}(\mathbf r,\omega)\bigr\rangle = 0,
\qquad
\bigl\langle \delta a_i^{P}(\mathbf r_0,\omega)\,\xi_j^{SV*}(\mathbf r,\omega)\bigr\rangle = 0,
\label{eq:incoherence_assumption}
\end{equation}
and similarly for the conjugate cross-terms. Consequently, second-order
correlations contain the power weights \(w_P\) and \(w_{SV}\), rather than
their square roots. The diffuse-field prescription used to assign these
weights, and its relation to the full-space P--S mixture, are derived in
App.~\ref{app:modal_fractions}.

\subsubsection{Mixed Correlation Tensors}

Because the mixed cross-terms evaluate to zero, any quadratic correlation tensor of the mixed field reduces to a simple weighted sum of the independent modal tensors. This holds for both the displacement-correlation tensor $\mathbf{C}$ and the gravitoelastic tensor $\mathbf{c}$:
\begin{align}
\mathbf{C}^{\mathrm{mix}}(\mathbf r',\mathbf r) &= w_P\,\mathbf{C}^{P}(\mathbf r',\mathbf r) + w_{SV}\,\mathbf{C}^{SV}(\mathbf r',\mathbf r), \label{eq:C_mixed_general}\\
\mathbf{c}^{\mathrm{mix}}(\mathbf r,\omega) &= w_P\,\mathbf{c}^{P}(\mathbf r,\omega) + w_{SV}\,\mathbf{c}^{SV}(\mathbf r,\omega). \label{eq:c_mixed_general}
\end{align}

These linear superposition formulas can be applied to two distinct theoretical models depending on the required boundary conditions:
\begin{itemize}
    \item \textbf{Formal Single-Wave Mixture:} Built from the pure, single upward-propagating $P$ and $SV$ waves. This is a formal mixture of analytic building blocks, useful for isolated code checks and theoretical analysis, but it does not physically satisfy the free-surface boundary conditions.
    \item \textbf{Physical Incident-Wave Mixture:} Built from the full, free-surface-consistent half-space solutions ($P \to P + P_{\mathrm{refl}} + SV_{\mathrm{conv}}$ and $SV \to SV + SV_{\mathrm{refl}} + P_{\mathrm{conv}}$). This is the correct physical model for half-space validation, as the seismic field must satisfy the surface boundary conditions regardless of whether the observer is above or below ground.
\end{itemize}

\subsubsection{Integral Consistency}

This incoherence framework introduces no new angular integrals and no new Newtonian kernels. Because the gravitoelastic integral operator is linear, substituting Eq.~\eqref{eq:C_mixed_general} into the gravitoelastic integral immediately returns Eq.~\eqref{eq:c_mixed_general}. 

For instance, an underground test mass evaluates the exact interior relation:
\begin{equation}
c_{ij}^{\mathrm{mix}}(\mathbf r,\omega)
=
G\rho_0\,\mathrm{PV}\!\int_V d^3r'\,
T_{i\alpha}(\mathbf r',\mathbf r_0)\,
C_{\alpha j}^{\mathrm{mix}}(\mathbf r',\mathbf r)
+
\frac{4\pi}{3}G\rho_0\,C_{ij}^{\mathrm{mix}}(\mathbf r_0,\mathbf r).
\label{eq:c_mix_interior_integral}
\end{equation}
The weighted sum passes directly through the principal value integral and the distributional local term, proving that the closed-form solutions and the integral representations remain perfectly consistent for multimodal fields.

\subsection{Test mass with spherical cavity}

For both Kernel-contraction and bulk-surface models, the leading-order cavity-corrected closed form is the same:

\begin{equation}
\label{eq:closed-form-cavity}
  c_{ij}^{\mathrm{cf,cav}}(\mathbf r,\omega)
  = c_{ij}(\mathbf r,\omega)
  - \frac{4\pi}{3}G\rho_0\,C_{ij}(\mathbf r_0,\mathbf r;\omega).
\end{equation}
Here:
$c_{ij}(\mathbf r,\omega)$ is the exact no-cavity closed form, and $C_{ij}(\mathbf r_0,\mathbf r;\omega)$ is obtained from the displacement tensor by setting the source point to the cavity center. Eq.~\eqref{eq:closed-form-cavity} is a small-cavity asymptotic reference. It is not an exact finite-cavity closed form. Also, Eq.~\eqref{eq:closed-form-cavity} is valid for Rayleigh waves and the full-space medium.

\subsection{Above-ground vs underground test mass}

Changing the test-mass position modifies only the gravitational coupling, not the seismic field. Therefore, the displacement correlation tensor \(C_{\alpha j}\) and the derived seismic quantities are identical for the underground and above-ground cases. In contrast, the gravitoelastic correlation changes because the gravitational potential, and hence the induced acceleration, depend on the test-mass position.

\section{Homogeneous isotropic full-space medium}
\label{sec:body_full}
In a homogeneous full space, compressional (P) waves generate NN through volumetric density perturbations in the bulk of the medium, whereas shear (S) waves contribute only in the presence of boundaries, such as cavity walls. These contributions are particularly relevant for underground detectors, where the seismic wave field may contain a significant body-wave component. Because the homogeneous full-space problem is translationally invariant, we choose the test mass as the coordinate origin, \(\mathbf r_0=\mathbf0\); \(\mathbf r_0\) is retained in the analytical formulas only to display the relative geometry.

The full-space displacement-correlation kernels and the exact homogeneous,
principal-value, and finite-cavity gravitoelastic tensors are collected in
App.~\ref{ssec:bw_models}. Here we establish the relation between the
finite-cavity and principal-value representations used in the numerical
validation.

\subsection{Approximate representations of the cavity gravitoelastic tensor}

Eq.~\eqref{eq:kc-cavity} is the exact definition of the integral of the cavity gravitoelastic tensor, while Eq.~\eqref{eq:bw_mix_cav} is the exact closed-form representation. In the small-cavity limit, the finite-cavity closed form reduces to the principal-value approximation \cite{Harms2019}. Comparisons labelled ``approximate cavity'' deliberately compare the numerical integral with a finite excluded source volume of radius \(a\) against this principal-value analytical approximation. Their reported discrepancy therefore contains both numerical-integration error and the finite-radius error of the principal-value approximation. \\
Eq.~\eqref{eq:closed-form-cavity} can also be used to derive an approximate closed-form expression for Newtonian noise generated by full-space body waves in the presence of an underground cavity. Importantly, this representation must explicitly account for S-waves; although S-waves produce no bulk density changes in a uniform full space, their displacement of the cavity walls acts as a moving surface mass, making them a significant contributor to the total cavity Newtonian noise.

\paragraph{Relation to the Principal-Value Model.}

The principal-value (PV) model is recovered in the limit of vanishing cavity radius. Using the small-argument expansion of the spherical Bessel function, \(
\frac{j_1(x)}{x}
=
\frac{1}{3}-\frac{x^2}{30}+O(x^4),
\)
the cavity multipliers expand as:
\begin{align}
M_P^{\mathrm{cavity}}
=
8\pi\frac{j_1(k_P a)}{k_P a}
=
\frac{8\pi}{3}
-\frac{8\pi}{30}(k_P a)^2
+O\!\left((k_Pa)^4\right),
\nonumber
\\
M_S^{\mathrm{cavity}}
=
-4\pi\frac{j_1(k_S a)}{k_S a}
=
-\frac{4\pi}{3}
+\frac{4\pi}{30}(k_S a)^2
+O\!\left((k_Sa)^4\right),
\end{align}

Taking the zero-radius limit $(a\to0)$ isolates the leading terms, which exactly match the principal-value multipliers:
\begin{equation}
M_P^{\mathrm{PV}}=\frac{8\pi}{3},
\qquad
M_S^{\mathrm{PV}}=-\frac{4\pi}{3}.
\end{equation}

Consequently, the total gravitoelastic tensor reduces to the PV form:
\begin{equation}
\lim_{a\to0}
c_{ij}^{\mathrm{cavity}}(\mathbf r,\omega)
=
c_{ij}^{\mathrm{PV}}(\mathbf r,\omega).
\end{equation}

Thus, the principal-value formulation is the zero-radius limit of the exact cavity model. For sufficiently small cavity radius \(a\), the cavity gravitoelastic tensor may therefore be accurately approximated by the PV closed-form expression.

\section{Numerical Validation and Convergence Studies}
\label{sec:validation}
In this section, we validate the numerical integral formulations against the analytical closed-form models. We verified all considered configurations, including real and imaginary tensor components and both integral formulations. For conciseness, we present a representative subset of validation results.

\subsection{Setup}
Unless explicitly varied in the frequency and cavity-radius studies, the seismic frequency is \(f=5\,\mathrm{Hz}\) and the cavity radius is \(a=40\,\mathrm{m}\). All simulations use a Poisson ratio \(\nu=0.27\), shear-wave speed \(\beta=2000\,\mathrm{m\,s^{-1}}\), and medium density \(\rho_0=2750\,\mathrm{kg\,m^{-3}}\). The corresponding P-wave speed is obtained from the isotropic elastic relation \(\alpha/\beta=\sqrt{2(1-\nu)/(1-2\nu)}\), while the Rayleigh-wave speed is obtained from the physical sub-shear root of the secular equation given in Eq.~(40) of \cite{Harms2019}. All lengths are expressed relative to the reference wavelength \(\lambda_{\mathrm{ref}}\), where \(\lambda_{\mathrm{ref}} = \lambda_{\mathrm{R}}\) for Rayleigh waves, \(\lambda_{\mathrm{ref}} = \lambda_{\mathrm{P}}\) for P and P--S mixed waves, and \(\lambda_{\mathrm{ref}} = \lambda_{\mathrm{S}}\) for S-waves. For these material parameters, the diffuse-field prescription derived in App.~\ref{app:modal_fractions} gives \(p_{\mathrm{HS}}=0.150\) and \(p_{\mathrm{FS}}=0.081\). Accordingly, the half-space model uses \(15.0\%\) incident P and \(85.0\%\)
incident SV power within its restricted P--SV ensemble. The full-space model
uses \(8.1\%\) P and \(91.9\%\) combined-S one-axis spectral power, with
\(S_n(\omega)=S_P+S_S=1\). These fractions are fixed by the common
equipartition-based modal-density prescription and are not selected by minimizing the numerical
error. The Rayleigh and half-space calculations use unit displacement-amplitude
normalization for each incident modal population. The exact homogeneous
full-space benchmark is evaluated as a pure P-wave case, whereas
\(p_{\mathrm{FS}}=0.081\) applies to the mixed principal-value and
approximate-cavity configurations. \\
For half-space configurations, the test mass is at \(z_0=\pm h\) relative to the free surface, with \(h=250\,\mathrm{m}\). In full space, \(\mathbf r_0=\mathbf0\), and the finite source box is centered on the origin; varying \(L_z\) moves both vertical boundaries symmetrically without moving the test mass or observation points. A closed spherical underground cavity must remain entirely inside the elastic half-space, so cavity configurations are evaluated only when \(h>a\). Consequently, the zero-depth sample is omitted from underground-cavity depth sweeps, while the remaining half-space cases retain the complete sampled range. \\
Map evaluations use a horizontal observation slice at \(z=-100\,\mathrm{m}\) relative to the half-space surface and at \(z=+150\,\mathrm{m}\) relative to the full-space test mass. The full-space value translates the former convention \(\mathbf r_0=(0,0,-250\,\mathrm{m})\), \(z=-100\,\mathrm{m}\), and therefore preserves its \(150\,\mathrm{m}\) separation. The square has half-length \(1.5\,\lambda_{\mathrm{ref}}\). The field is evaluated on a \(17\times17\) grid and interpolated onto a \(101\times101\) grid; convergence studies use a \(9\times9\) grid. \\
To validate the numerical models against the closed-form gravitoelastic correlation tensors, we performed three primary convergence tests:
\begin{enumerate}
    \item \textbf{Discretization and Truncation Sweep:} We evaluated the spatial resolution (expressed as points per wavelength, \(\mathrm{ppw}\)), the horizontal domain size (\(L_{xy}\)), and the vertical truncation length (\(L_z\)) for all models. The test-mass depth or height (\(h\)) was varied only for half-space models, since \(h\) is undefined in homogeneous full space. In each test, one parameter was varied while the others were fixed (Table~\ref{tab:convergence_parameters}).
    
    \item \textbf{Cavity-Radius Convergence:} Maintaining all other parameters constant, we studied how the error changed with the normalized cavity radius across the values \(a/\lambda_{\mathrm{ref}} \in \{0.3, 0.2, 0.1, 0.08, 0.06, 0.04, 0.02, 0.01, 0.005\}\).
    
    \item \textbf{Frequency Convergence:} We examined how the error scales with frequency \(f\) using 179 unique values: 30 linearly spaced points from \(0.01\) to \(0.1\,\mathrm{Hz}\) and 150 linearly spaced points from \(0.1\) to \(20\,\mathrm{Hz}\), with the shared \(0.1\,\mathrm{Hz}\) point retained only once. The requested physical grid has \(L_{xy}^{\mathrm{req}}=2400\,\mathrm{m}\), \(L_z^{\mathrm{req}}=1200\,\mathrm{m}\), \(N_{xy}^{\mathrm{req}}=120\), and \(N_z^{\mathrm{req}}=60\). For underground half-space cases, the spacing, counts, and realized extents are reconstructed according to the depth-alignment rule in App.~\ref{sec:geometry}.
\end{enumerate}

\begin{table}[htbp]
\centering
\caption{Reference configurations and parameter sweeps. In half space, \(L_z\) is the depth and maps use \(h=250\,\mathrm{m}\); in full space, \(L_z\) is the total box height centered on \(\mathbf r_0=\mathbf0\). The \(h/\lambda_{\rm ref}\) sweep is half-space only. For underground-cavity cases, only samples satisfying \(h>a\) are used; thus the \(h/\lambda_{\rm ref}=0\) point is omitted from those curves.}
\label{tab:convergence_parameters}
\begin{tabular}{lccc}
\hline
Parameter & Rayleigh Ref. & Body-Wave Ref. & Sweep Values \\
\hline
\(\mathrm{ppw}\) 
& \(30\) 
& \(20\) 
& \(\{2, 4, 8, 12, 16, 20, 24, 28, 32, 36, 40\}\) \\

\(L_{xy}/\lambda_{\mathrm{ref}}\) 
& \(10\) 
& \(8\) 
& \(\{1, 2, 4, 6, 8, 10, 12\}\) \\

\(L_z/\lambda_{\mathrm{ref}}\) 
& \(5\) 
& \(8\) 
& \(\{1, 2, 4, 6, 8, 10\}\) \\

Half-space \(h\)
& \(250\,\mathrm{m}\) 
& \(250\,\mathrm{m}\) 
& \(\{0, 0.25, 0.5, 0.75, 1.0, 1.25, 1.5, 1.75, 2.0\} \times \lambda_{\mathrm{ref}}\) \\
\hline
\end{tabular}
\end{table}

\subsection{Results}
The results presented below cover a range of wave types, geometries, and numerical formulations. To keep the figure captions and discussion concise, we use the abbreviations summarized in Table~\ref{tab:abbreviations}.

\begin{table}[htbp]
    \centering
    \caption{Abbreviations used throughout the validation results to denote wave types, geometries, and numerical formulations.}
    \label{tab:abbreviations}
    \small
    \begin{tabular}{@{}llll@{}}
        \hline
        \textbf{Abbreviation} & \textbf{Definition} & \textbf{Abbreviation} & \textbf{Definition} \\
        \hline
        \textbf{R} & Rayleigh waves & \textbf{B} & Body waves \\
        \textbf{HS} & Half-space & \textbf{FS} & Full space \\
        \textbf{AG} & Above-ground test mass & \textbf{UG} & Underground test mass \\
        \textbf{Cav} & Cavity configuration & \textbf{BS} & Bulk-surface formulation \\
        \textbf{KC} & Kernel-contraction formulation & \textbf{PV} & Full-space PV formulation \\
        \textbf{App Cav} & Approximate cavity model & \textbf{Mix} & Full-space P/S mixture \\
        \textbf{Inc Mix} & Half-space incident P/SV mixture & & \\
        \hline
    \end{tabular}
\end{table}

\subsubsection{General Convergence Analysis}
\begin{figure}[h]
\centering
\includegraphics[width=\textwidth]{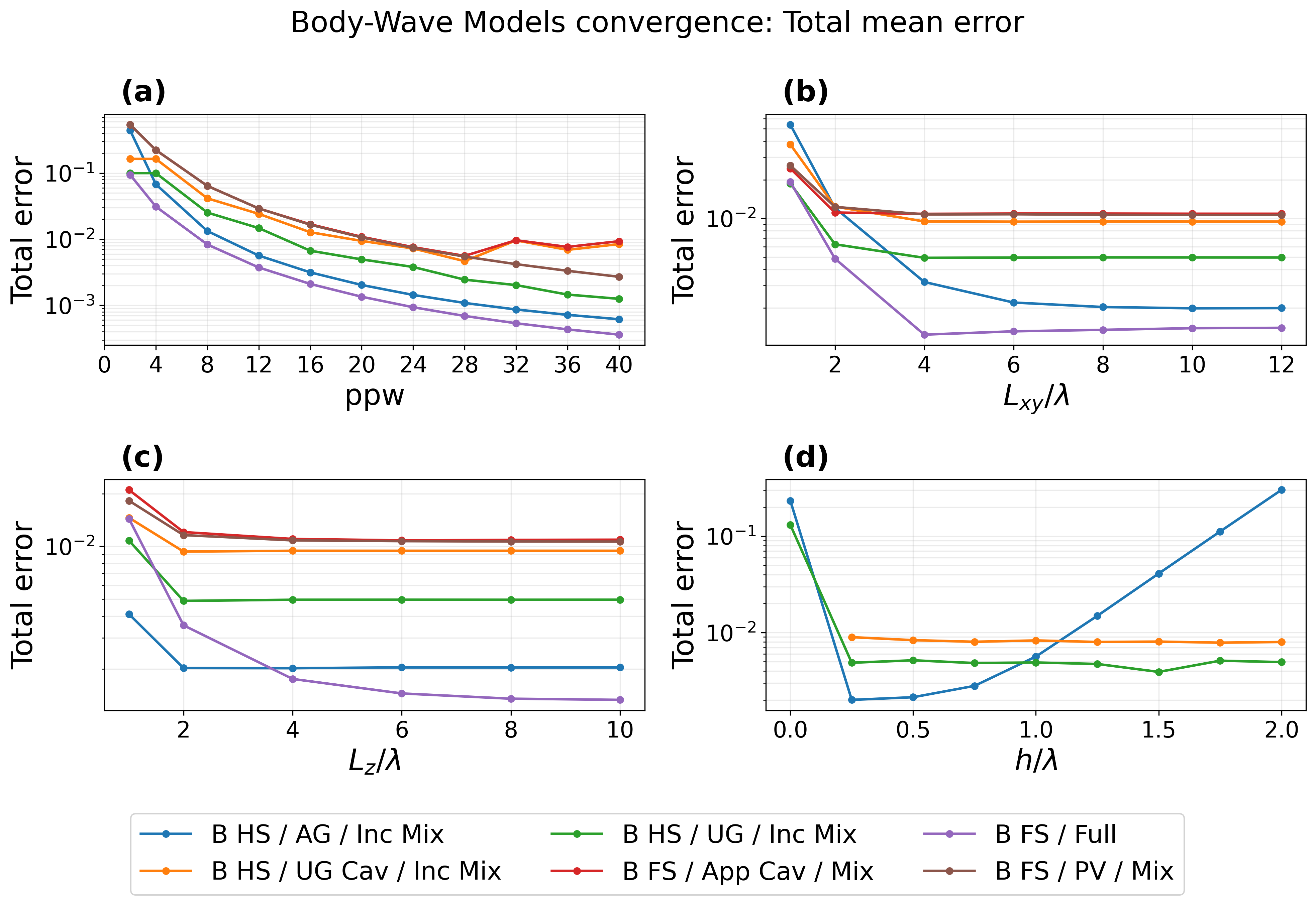}
\caption{Convergence tests for the body-wave models as functions of: (a) spatial resolution \(\mathrm{ppw}\), (b) horizontal domain size \(L_{xy}/\lambda_{\mathrm{ref}}\), (c) vertical truncation length \(L_z/\lambda_{\mathrm{ref}}\), and (d) half-space test-mass position at a fixed observation slice. Full-space cases are omitted from panel (d). The zero-depth point is excluded for underground cavity configurations to maintain the cavity below the free surface.}
\label{fig:body_convergence}
\end{figure}

\begin{figure}[h]
\centering
\includegraphics[width=\textwidth]{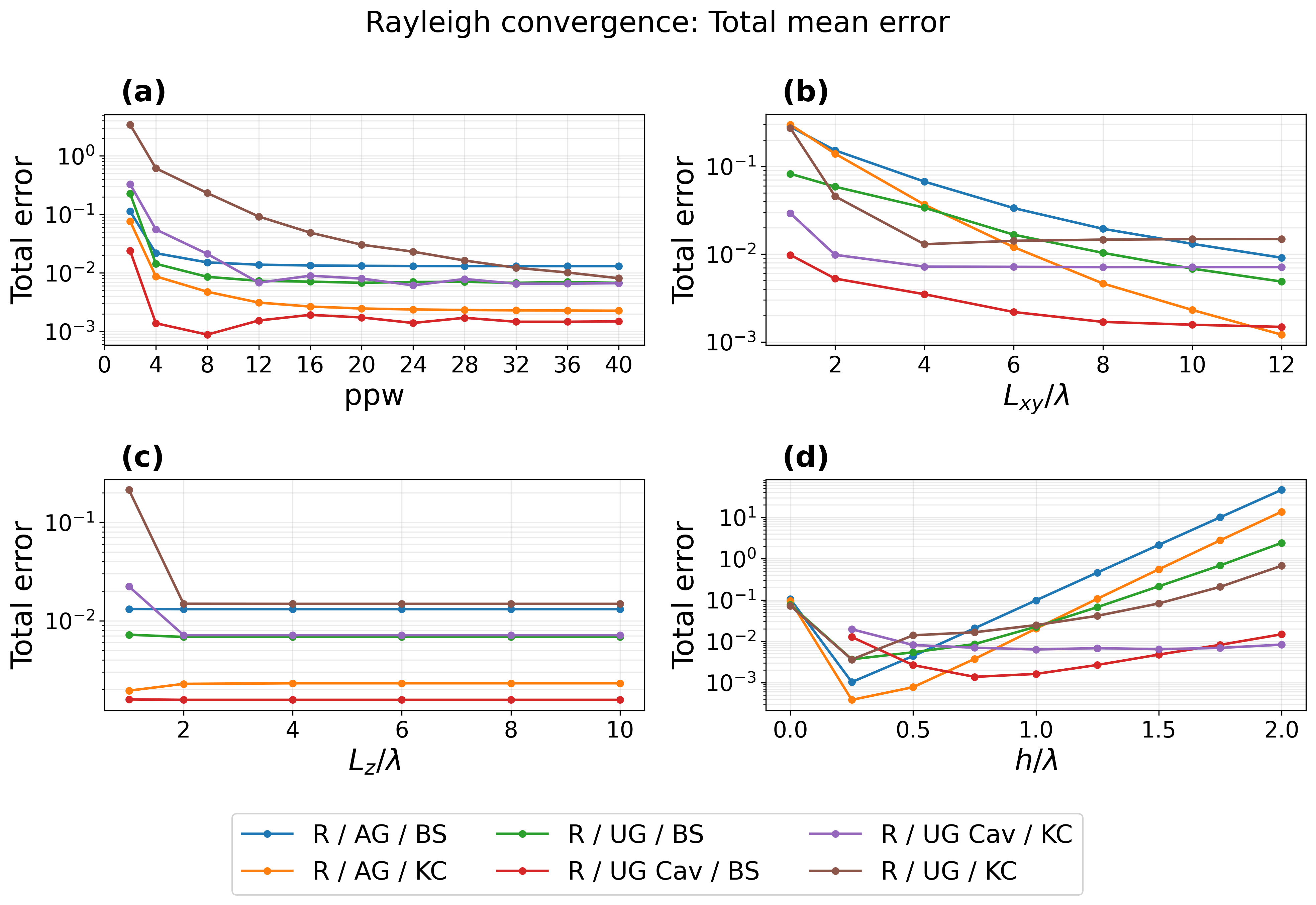}
\caption{Convergence tests for the Rayleigh-wave models as functions of: (a) spatial resolution \(\mathrm{ppw}\), (b) horizontal domain size \(L_{xy}/\lambda_{\mathrm{ref}}\), (c) vertical truncation length \(L_z/\lambda_{\mathrm{ref}}\), and (d) test-mass position at a fixed observation slice. The zero-depth point is excluded for underground cavity configurations.}
\label{fig:rayleigh_convergence}
\end{figure}

Figures~\ref{fig:body_convergence} and \ref{fig:rayleigh_convergence} show the convergence of the body-wave and Rayleigh-wave models with spatial resolution, horizontal and vertical domain size, and test-mass position. The points-per-wavelength curves predominantly decrease smoothly toward model-dependent residual levels. No sustained oscillatory behavior is observed. A few cavity cases show isolated non-monotonic deviations, including a local increase near \(\mathrm{ppw}=32\), but the overall convergence remains clear.

At \(\mathrm{ppw}=40\), the mean total error ranges from \(6.17\times10^{-4}\) to \(8.44\times10^{-3}\) for the half-space body-wave models and from \(3.61\times10^{-4}\) to \(9.34\times10^{-3}\) for the full-space body-wave models. The corresponding Rayleigh-wave errors range from \(1.49\times10^{-3}\) to \(1.31\times10^{-2}\). The smallest body-wave error is obtained for the exact homogeneous full-space model, whereas the larger residuals generally occur in configurations containing a cavity correction or a free surface.

Increasing the horizontal and vertical integration extents initially reduces the truncation error, after which the curves approach stable, model-dependent plateaus.

The test-mass-position sweeps are geometry-sensitivity studies rather than strict convergence tests because changing the test-mass position also changes the physical response. Full-space models are therefore omitted from these sweeps because the test mass is fixed at the centre of the truncated full-space domain. In underground cavity cases, the zero-depth point is excluded to preserve the required condition that the cavity remains below the free surface.

\begin{figure}[h]
\centering
\includegraphics[width=\textwidth]{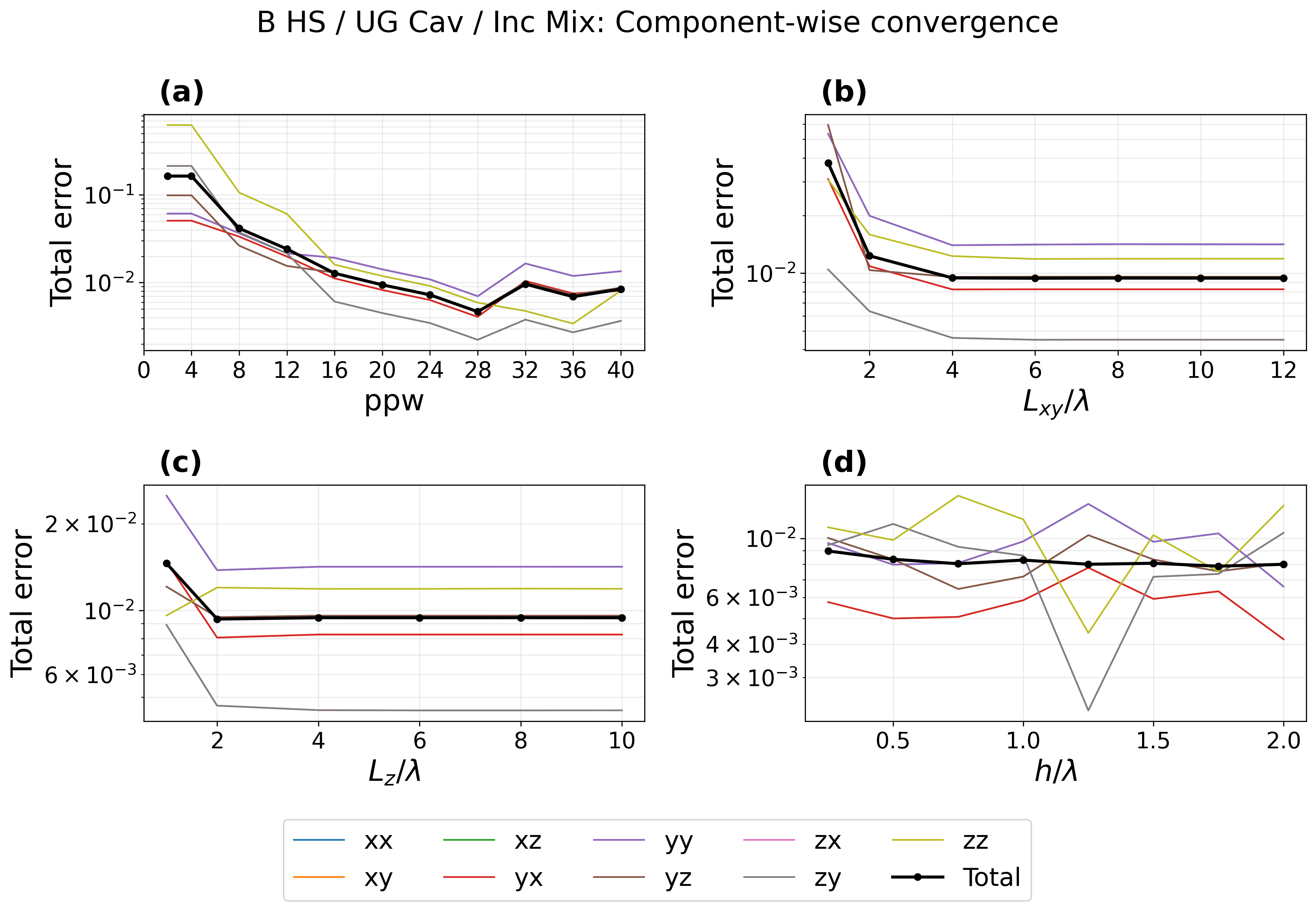}
\caption{Component-wise and total convergence errors for the half-space underground-cavity mixed P--SV body-wave model as functions of: (a) spatial resolution \(\mathrm{ppw}\), (b) horizontal domain size \(L_{xy}/\lambda_{\mathrm{ref}}\), (c) vertical truncation length \(L_z/\lambda_{\mathrm{ref}}\), and (d) test-mass depth at a fixed observation slice. The zero-depth point is excluded because the underground cavity must remain below the free surface.}
\label{fig:component_convergence}
\end{figure}

\begin{figure}[h]
\centering
\includegraphics[width=\textwidth]{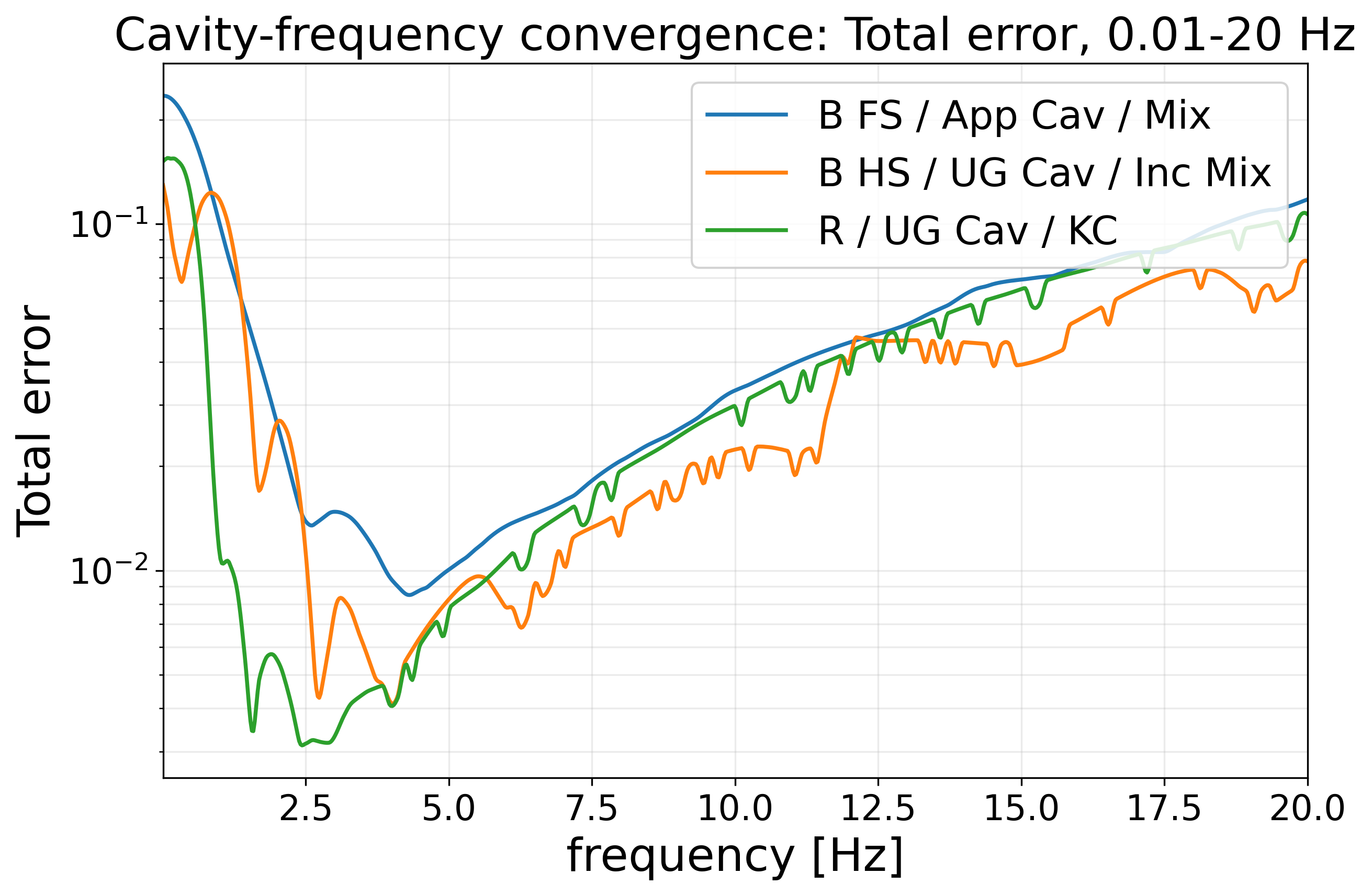}
\caption{Frequency dependence of the total error over \(0.01\)--\(20\,\mathrm{Hz}\) for the three representative cavity configurations. The curves exhibit broad, model-dependent minima at intermediate frequencies, with local variations arising from the combined effects of truncation and spatial resolution.}
\label{fig:freq_convergence}
\end{figure}

\begin{figure}[h]
\centering
\includegraphics[width=\textwidth]{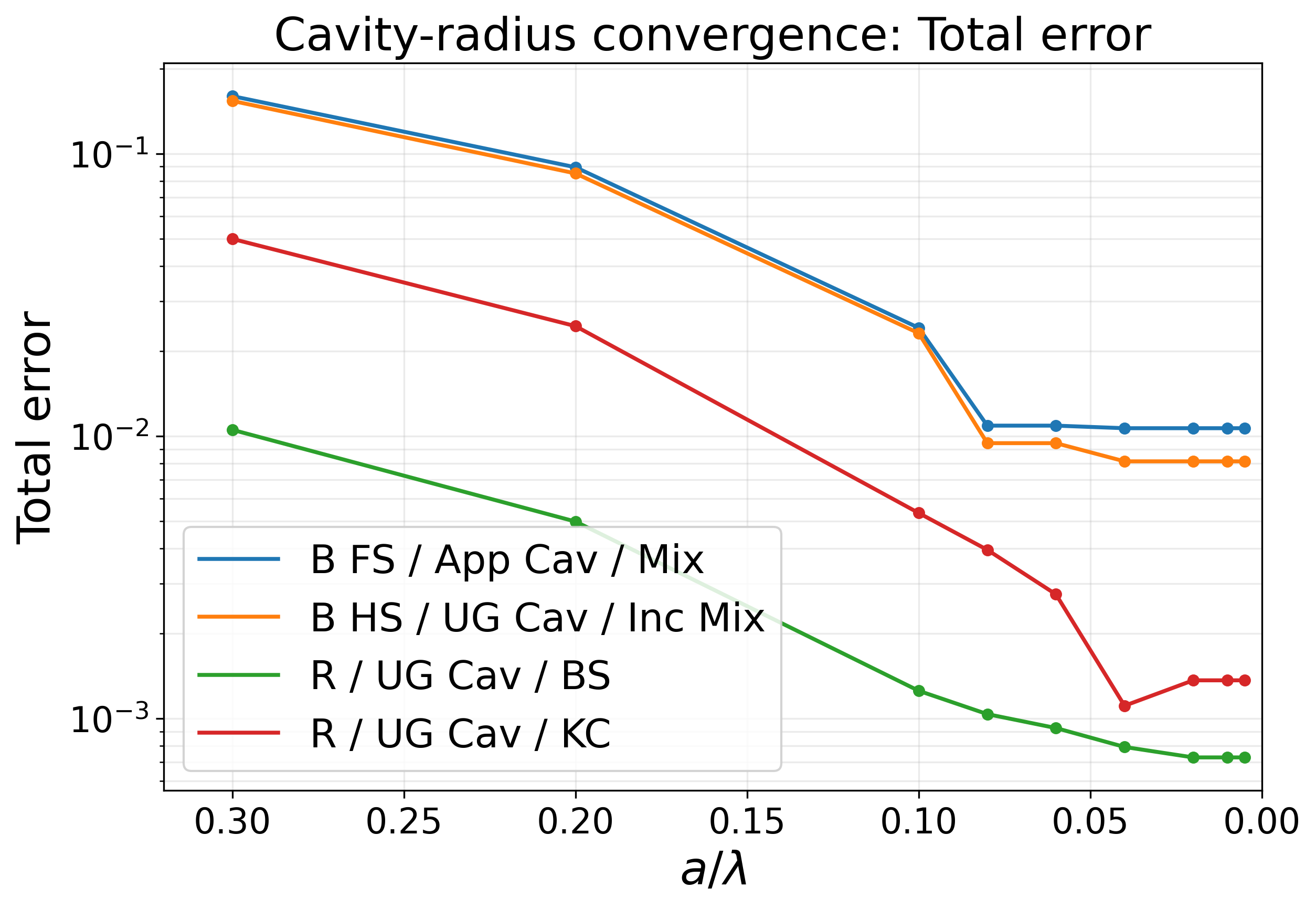}
\caption{Total error as a function of normalized cavity radius. The body-wave models and Rayleigh bulk-surface formulation approach small-radius plateaus, while the Rayleigh kernel-contraction formulation exhibits a shallow intermediate minimum before approaching its small-radius plateau.}
\label{fig:cavity_radius_convergence}
\end{figure}

\subsubsection{Component-Level Convergence}
Figure~\ref{fig:component_convergence} resolves the convergence error into the nine tensor components for the half-space underground-cavity mixed P--SV model. At coarse resolutions, particularly \(\mathrm{ppw}=2\) and \(4\), the \(zz\) component provides the largest contribution to the error, followed by the \(zx\) and \(zy\) components. At \(\mathrm{ppw}\geq16\), the remaining error is generally dominated by the \(xx\) and \(yy\) components. The isolated increase near \(\mathrm{ppw}=32\) is also associated mainly with these horizontal diagonal components.

After the horizontal and vertical domain-size curves reach their plateaus, the principal residuals remain in the \(xx\) and \(yy\) components, followed by \(zz\). The mean total error follows the overall component-wise behavior and therefore provides a useful scalar measure of convergence, while the individual curves identify which tensor components control the residual in each numerical regime.

\subsubsection{Frequency and Cavity-Radius Convergence}
Figure~\ref{fig:freq_convergence} shows the frequency dependence over \(0.01\)--\(20\,\mathrm{Hz}\). Each model exhibits a broad, model-dependent minimum at an intermediate frequency, with local variations superimposed on the overall trend. The minima occur near \(2.50\,\mathrm{Hz}\) for the Rayleigh underground-cavity model, \(3.97\,\mathrm{Hz}\) for the half-space body-wave cavity model, and \(4.37\,\mathrm{Hz}\) for the full-space approximate-cavity model. The corresponding minimum errors are \(3.17\times10^{-3}\), \(4.18\times10^{-3}\), and \(8.55\times10^{-3}\), respectively.

Figure~\ref{fig:cavity_radius_convergence} shows the dependence on the normalized cavity radius. The half-space and full-space body-wave errors approach small-radius plateaus of approximately \(8.14\times10^{-3}\) and \(1.07\times10^{-2}\), respectively, for \(a/\lambda_{\mathrm{ref}}\leq0.04\). The Rayleigh bulk-surface formulation approaches a plateau of approximately \(7.28\times10^{-4}\) for \(a/\lambda_{\mathrm{ref}}\leq0.02\). The Rayleigh kernel-contraction formulation reaches a shallow minimum of approximately \(1.11\times10^{-3}\) near \(a/\lambda_{\mathrm{ref}}=0.04\) and then approaches approximately \(1.37\times10^{-3}\) at smaller radii.

\subsubsection{Maps}
\begin{figure}[h]
\centering
\includegraphics[width=\textwidth]{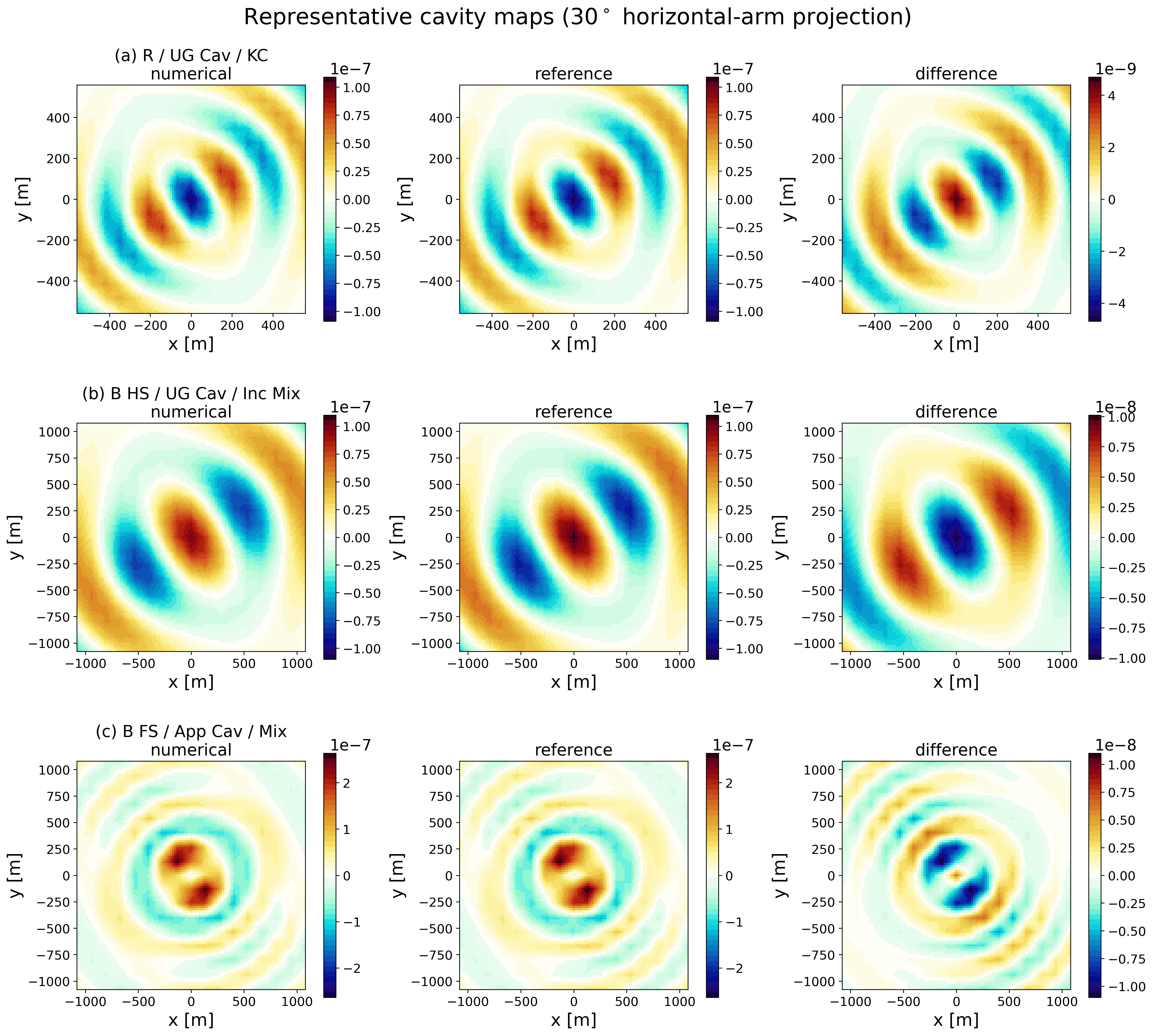}
\caption{Representative comparisons of the real gravitoelastic correlation projected along a horizontal detector arm oriented at \(30^\circ\) to the \(x\)-axis. The rows show: (a) the Rayleigh underground-cavity model, (b) the half-space body-wave underground-cavity mixed P--SV model, and (c) the full-space body-wave approximate-cavity model. Each row contains the numerical field, analytical reference, and their difference.}
\label{fig:representative_maps}
\end{figure}

Figure~\ref{fig:representative_maps} compares the numerical and analytical real parts of the gravitoelastic tensor projected along a horizontal detector arm oriented at \(30^\circ\) to the \(x\)-axis. The three rows represent the Rayleigh underground-cavity, half-space body-wave underground-cavity, and full-space body-wave approximate-cavity configurations.

The numerical maps reproduce the symmetry, sign changes, spatial lobes, and amplitudes of the analytical fields. At each of the \(101\times101\) interpolated points, the map discrepancy is the absolute projected residual divided by the peak absolute analytical projected field over the complete map. The mean and maximum of this quantity are \(0.85\%\) and \(4.31\%\) for the Rayleigh model, \(2.36\%\) and \(9.20\%\) for the half-space body-wave model, and \(0.53\%\) and \(4.17\%\) for the full-space model.

\section{Discussion}\label{sec:discussion}

The numerical validation demonstrates that the Cartesian midpoint-grid formulation reproduces the analytical gravitoelastic correlation tensors for Rayleigh waves and for body waves in half-space and full-space media. The tests include above-ground, underground, and cavity configurations. Accuracy is assessed using all nine tensor components, while the spatial comparisons use the tensor projected along a horizontal detector arm oriented at \(30^\circ\).

The points-per-wavelength curves predominantly decrease smoothly with increasing resolution. Their eventual plateaus are model dependent because the total error contains contributions from grid discretization, finite-domain truncation, and the treatment of local or cavity terms. The isolated non-monotonic deviations observed in a few cavity cases do not constitute sustained oscillatory convergence. They arise when changes in grid spacing alter the balance among component contributions or the Cartesian representation of the spherical cavity boundary.

The horizontal and vertical domain-size studies show smoother behaviour. Once the integration domain contains the region that contributes appreciably to the gravitoelastic response, increasing its extent produces little further improvement. The resulting plateaus indicate that the selected domains are adequate for validating the numerical integrals without requiring an impractically large computational volume.

The test-mass-position sweeps show a U-shaped relative-error trend for the Rayleigh-wave and above-ground body-wave models. In these configurations, the response decreases with test-mass distance because of evanescent factors such as \(e^{-k_R h}\) and \(e^{-k_\rho h}\). At small distances, the gravitational kernel places greater weight on nearby source cells and near-surface contributions, making the numerical result more sensitive to spatial discretization. At intermediate distances, near-field discretization and response amplitude are most favourably balanced. At larger distances, the reference response becomes small, so even a modest absolute discrepancy produces a larger relative error. The underground body-wave models do not exhibit the same pronounced U-shaped behaviour because their propagating bulk contributions do not decay exponentially with depth. Their depth dependence is instead governed mainly by phase-dependent interference among incident, reflected, and mode-converted components.

The component-wise results show why a single total error cannot describe every aspect of convergence. The dominant component changes with spatial resolution: \(zz\) controls much of the coarse-grid error, whereas \(xx\) and \(yy\) dominate much of the residual at finer resolutions. The mean total error nevertheless follows the general component-wise convergence and is therefore suitable for comparing the numerical configurations, provided that representative component-level checks are also reported.

Mixing P and shear contributions can introduce physical cancellation in signed tensor components. When the two weighted contributions have opposite signs or phases, cancellation tends to become more pronounced as their magnitudes become comparable and weaker when either contribution dominates. The mixed component can consequently be much smaller than its separate modal contributions, causing modest absolute numerical errors to appear as a larger relative error.

This effect is weak in the no-cavity configurations considered here. The relevant P and SV contributions reinforce in the half-space models, while transverse waves do not contribute to the bulk gravitoelastic response of the exact homogeneous full-space model. Cancellation can become more apparent in cavity configurations because the cavity correction modifies the modal contributions differently and can give them opposite signs. For the adopted diffuse-field equipartition fractions, the shear contribution generally dominates and the cancellation remains limited. For comparison, sensitivity calculations using a half-space P-wave fraction of \(0.7\) produced more noticeable cancellation in horizontal tensor components because the weighted P and SV contributions became more comparable.

The frequency dependence results from two competing numerical limitations. At low frequency, the finite computational domain covers fewer wavelengths and truncation becomes more important. At high frequency, shorter wavelengths require finer spatial resolution. For the full-space approximate-cavity comparison, the high-frequency discrepancy also includes the reduced accuracy of the principal-value approximation as the cavity becomes larger relative to the wavelength. The broad minima at intermediate frequencies identify the range in which these effects are most favourably balanced for the selected grid.

The cavity-radius plateaus have a related interpretation. When the cavity becomes small compared with both the wavelength and the grid scale, reducing its radius further produces little change in the measured numerical error. The plateau therefore reflects the combined small-cavity and finite-resolution limits rather than a loss of physical dependence on cavity size.

The projected maps provide a direct spatial validation complementary to the convergence curves. The numerical fields reproduce the analytical symmetries and lobe structures even where local residuals are larger. Peak normalization is used because pointwise relative errors become ill-conditioned near zero crossings of the reference field.

The cavity models used in this validation describe the leading-order gravitational effect of excluding material around the test mass while retaining the prescribed unscattered seismic field. They do not include elastic scattering or additional mode conversion at a finite spherical cavity wall. The comparisons therefore validate the numerical implementation of the stated analytical model, rather than a complete finite-cavity elastodynamic solution. Incorporating cavity-wall scattering, heterogeneous material properties, and realistic underground geometry constitutes a separate extension of the physical model.

\section{Conclusion}
\label{sec:conclusion}
A Cartesian midpoint-grid framework has been developed and validated for evaluating gravitoelastic correlation tensors associated with Rayleigh waves and with body waves in half-space and full-space media. The comparison covers above-ground, underground, and cavity configurations, evaluates all nine tensor components, and includes spatial projections along a representative detector arm.

The convergence and spatial comparisons show that the numerical tensors reproduce the analytical symmetries, amplitudes, and spatial structures with high accuracy. The remaining model-dependent residuals are controlled by spatial discretization, finite-domain truncation, and the treatment of local and cavity terms. The frequency and cavity-radius studies identify the regimes in which these effects and the small-cavity principal-value approximation become important, while the component-wise analysis shows how signed modal cancellation can amplify relative errors.

Under the assumptions of homogeneous media, a flat free surface, isotropic modal ensembles, and an unscattered seismic field, the framework provides a controlled numerical foundation for extensions to heterogeneous media, realistic topography, anisotropic seismic fields, and finite-cavity elastic scattering.


\textbf{Data and code availability}\\
The numerical data supporting the findings of this study and the source code used to generate them are not publicly available because a permanent repository has not yet been established. They are available from the corresponding author upon reasonable request.

\begin{appendices}

\section{Geometry and Coordinate System}
\label{sec:geometry}

We use $\mathbf r=(\bm\rho,z)$, with $\bm\rho=(x,y)$ and \(z\) upward. For half-space models the origin is on the surface \(z=z_s(\bm\rho)\), \(z_s(\mathbf0)=0\), and the medium occupies \(z\le z_s\); the flat model has \(z_s=0\). The homogeneous full-space medium occupies \(\mathbb R^3\), has no surface, and is translationally invariant, so we choose the test mass as the coordinate origin. \\
The test mass is at $\mathbf r_0=(0,0,z_0)$. In the flat half space, \(z_0=-h\) and \(+h\) denote underground and above-ground positions, respectively. In full space, \(\mathbf r_0=\mathbf0\) and \(h\) is not defined. \\
With the TM on the vertical axis, the Euclidean distance between a source point and the test mass simplifies to:
\begin{equation}
    |\mathbf{r}' - \mathbf{r}_0| = \sqrt{\rho'^2 + (z' - z_0)^2},
\end{equation}
where $\rho' = |\bm{\rho}'| = \sqrt{x'^2 + y'^2}$ is the horizontal distance to the source, and its absolute distance to the coordinate origin is $r' = |\mathbf{r}'| = \sqrt{\rho'^2 + z'^2}$.

\section{Grid discretization}

We discretize the continuous source region using a structured Cartesian grid. Volume and surface integrals are evaluated via the midpoint rule, where each point represents the geometric center of a cell with volume $dV = \Delta x\,\Delta y\,\Delta z$ or an elementary surface patch with area $dS = \Delta x\,\Delta y$:
\begin{equation}
\int_V f(\mathbf r')\,d^3r' \approx \sum_{i,j,k} f(\mathbf r'_{ijk})\,dV, \qquad \int_S F(\bm{\rho}',0)\,d^2\rho' \approx \sum_{i,j} F(\bm{\rho}'_{ij},0)\,dS.
\end{equation}

The requested finite computational domain spans a horizontal area \((L_{xy}^{\mathrm{req}})^2\) centered on the test-mass vertical axis and has requested vertical extent \(L_z^{\mathrm{req}}\). If \(N_{\mathrm{ppw}}\) denotes the number of cell-centered source-grid points per reference wavelength, the target cubic spacing is $\Delta_{\mathrm{t}}=\frac{\lambda_{\mathrm{ref}}}{N_{\mathrm{ppw}}}$.

For an above-ground half-space test mass we set \(\Delta=\Delta_{\mathrm{t}}\). For an underground test mass at depth \(h\), we preserve the physical depth and impose exact depth alignment through
\begin{equation}
N_h=\max\!\left[1,\operatorname{round}\!\left(\frac{h}{\Delta_{\mathrm{t}}}\right)\right],
\qquad
\Delta=\frac{h}{N_h}.
\end{equation}
The cell counts are then reconstructed from the requested extents, \(N_{xy}=\operatorname{even}\{\operatorname{round}(L_{xy}^{\mathrm{req}}/\Delta)\}\) and \(N_z=\operatorname{round}(L_z^{\mathrm{req}}/\Delta)\), where \(\operatorname{even}\{\cdot\}\) raises an odd horizontal count to the next even integer. For full space, \(\Delta=\Delta_{\mathrm{t}}\) and both \(N_{xy}\) and \(N_z\) are made even so that the test mass remains at the center between cells. In every geometry,
\begin{equation}
\Delta x=\Delta y=\Delta z=\Delta,\qquad
L_{xy}=N_{xy}\Delta,\qquad L_z=N_z\Delta .
\end{equation}
Thus the realized counts, extents, and, after underground depth alignment, the realized PPW \(N_{\mathrm{ppw}}^{\mathrm{real}}=\lambda_{\mathrm{ref}}/\Delta\) can differ slightly from their requested values; both requested and realized quantities are recorded. The midpoint source coordinates $\mathbf r'_{ijk}=(x'_i,y'_j,z'_k)$ are
\begin{equation}
\begin{aligned}
x'_i&=\left(i+\frac{1}{2}-\frac{N_{xy}}{2}\right)\Delta,
&y'_j&=\left(j+\frac{1}{2}-\frac{N_{xy}}{2}\right)\Delta,\\
z'_k&=
\begin{cases}
-\left(k+\frac{1}{2}\right)\Delta,
&\text{half space},\\[1mm]
\left(k+\frac{1}{2}-\frac{N_z}{2}\right)\Delta,
&\text{full space}.
\end{cases}
\end{aligned}
\end{equation}
Here $i,j=0,\ldots,N_{xy}-1$ and $k=0,\ldots,N_z-1$. In half space, \(L_z\) is the realized depth and \(-L_z+\Delta/2\le z'_k\le-\Delta/2\); the first source layer is therefore centered at \(z=-\Delta/2\). For an underground test mass, \(h=N_h\Delta\) places it exactly halfway between the adjacent layers at \(z=-h\pm\Delta/2\). In full space, \(L_z\) is the realized total box height and \(-L_z/2+\Delta/2\le z'_k\le L_z/2-\Delta/2\). Even \(N_{xy}\) and \(N_z\) place \(\mathbf r_0=\mathbf0\) halfway between the eight central midpoint cells; increasing \(L_z\) expands both boundaries symmetrically.

\section{Error metrics}

Let $\mathbf{A}^{(n)}$ and $\mathbf{B}^{(n)}$ denote the numerical and reference tensors at the $n$-th observation point, respectively. To evaluate relative errors while avoiding division by zero for vanishing fields, we define a normalization weight $w^{(n)}$ based on the Frobenius norm $\|\cdot\|_F$ of the reference tensor: $w^{(n)} = \|\mathbf{B}^{(n)}\|_F$ if $\|\mathbf{B}^{(n)}\|_F \ge 10^{-15}$, and $w^{(n)} = 1$ otherwise. \\
The relative Frobenius error and the component-wise error for a single tensor pair are defined as:
\begin{equation}
\epsilon_{\mathrm{F}}^{(n)} = \frac{\|\mathbf{A}^{(n)} - \mathbf{B}^{(n)}\|_F}{w^{(n)}}, \qquad \epsilon_{ij}^{(n)} = \frac{|A_{ij}^{(n)} - B_{ij}^{(n)}|}{w^{(n)}}.
\end{equation}

The mean Frobenius and component-wise errors over all $N$ observation points are respectively given by $\overline{\epsilon}_{\mathrm{F}} = \frac{1}{N}\sum_{n=1}^N \epsilon_{\mathrm{F}}^{(n)}$ and $\overline{\epsilon}_{ij} = \frac{1}{N}\sum_{n=1}^N \epsilon_{ij}^{(n)}$. Finally, the total mean component error is calculated as the average across all nine tensor components:
\begin{equation}
\overline{\epsilon}_{\mathrm{total}} = \frac{1}{9}\sum_{i,j \in \{x,y,z\}} \overline{\epsilon}_{ij}.
\end{equation}

\section{Unified Master Equations}
\label{app:master_equations}

For any horizontally isotropic seismic field, the displacement, two-point correlation, gravitational potential, and Newtonian acceleration can be entirely described by a base amplitude $s$, a horizontal wavevector $\mathbf{k}_h$, a horizontal wavenumber $\kappa$, and a set of depth-dependent profile functions.

\subsection{Master Displacement, potential and acceleration}

The universal displacement vector is expressed in terms of generalized horizontal and vertical displacement profiles, $D_h(z)$ and $D_z(z)$:
\begin{equation}
    \bm\xi(\mathbf{r},\omega) = s e^{i\mathbf{k}_h\cdot\boldsymbol\rho} \left[ D_h(z)\hat{\mathbf{e}}_h + D_z(z)\hat{\mathbf{e}}_z \right].
\end{equation}

For a test mass located at $\mathbf{r}_0 = (\boldsymbol{\rho}_0, z_0)$, the exact gravitational potential and acceleration generated by the field take the compact form:
\begin{equation}
    \delta\Phi(\mathbf{r}_0, \omega) = \mathcal{F}(\boldsymbol{\rho}_0) \, \mathcal{B}_{\mathrm{pot}}(z_0),
\end{equation}
\begin{equation}
    \delta\mathbf{a}(\mathbf{r}_0, \omega)
=
\mathcal{F}(\boldsymbol{\rho}_0)
\left[
A_h(z_0)\hat{\mathbf{e}}_h
+
A_z(z_0)\hat{\mathbf{e}}_z
\right],
\end{equation}
where the horizontal phase operator is defined as:
\begin{equation}
    \mathcal{F}(\boldsymbol{\rho}_0) = 2\pi G \rho_0 s \, e^{i\mathbf{k}_h \cdot \boldsymbol{\rho}_0},
\end{equation}
and the mode-dependent acceleration amplitudes are related to the potential and vertical gravity profiles by
\begin{equation}
    A_h(z_0)
    =
    -i \kappa \mathcal B_{\rm pot}(z_0),
    \qquad
    A_z(z_0)
    =
    \mathcal B_{\rm vert}(z_0).
    \label{eq:Arho_Az_Bpot_Bvert}
\end{equation}

and the vertical acceleration profile is strictly the negative derivative of the potential profile:
\begin{equation}
    \mathcal{B}_{\mathrm{vert}}(z_0) \equiv -\partial_{z_0}\mathcal{B}_{\mathrm{pot}}(z_0).
\end{equation}

All dependence on wave type, boundary interactions, and observation depth is completely contained within the parameter mapping of $D_h$, $D_z$, $\mathcal{B}_{\mathrm{pot}}$, and $\mathcal{B}_{\mathrm{vert}}$.

\label{sec:wave_parameters}

\subsection{Body waves in a homogeneous half-space}
\label{ssec:b_wave_parameters}

For body waves, the common horizontal wavevector is $\mathbf{k}_h = \mathbf{k}_\rho = k_\rho(\cos\theta,\sin\theta,0)$. The P- and S-wave wavenumbers are $k_P=\omega/\alpha$ and $k_S=\omega/\beta$. Their vertical components satisfy $k_z^P=\sqrt{k_P^2-k_\rho^2}$ and $k_z^S=\sqrt{k_S^2-k_\rho^2}$. The P and SV polarizations are $\hat{\mathbf e}_\pm^P=\frac{\mathbf k_\rho\pm k_z^P\hat{\mathbf e}_z}{k_P}$, and $\hat{\mathbf e}_\pm^{SV}
=
\frac{\pm k_z^S\hat{\mathbf e}_\rho-k_\rho\hat{\mathbf e}_z}{k_S}$,
with $\hat{\mathbf e}_\pm^{SV}\cdot\mathbf k_\pm^S=0$.

\subsubsection{Displacement Mapping}

Because body-wave horizontal and vertical profiles share the same phase baseline, mapping them to the Master Equations directly preserves the imaginary unit $i$ in the cross-correlation terms of Eq.~\eqref{eq:C_general}. \\
An upward-incident P wave produces reflected P and SV waves at the free surface \cite{AkiRichards2002}. The generalized parameters map to $s = \xi_0^P$, $D_h(z) = U_\rho^{(P)}(z)$, and $D_z(z) = U_z^{(P)}(z)$, yielding:
\begin{align}
    U_\rho^{(P)}(z) &= \frac{k_\rho}{k_P}e^{ik_z^P z} + R_{PP}\frac{k_\rho}{k_P}e^{-ik_z^P z} - R_{PS}\frac{k_z^S}{k_S}e^{-ik_z^S z}, \label{eq:u_rho_p}\\
    U_z^{(P)}(z) &= \frac{k_z^P}{k_P}e^{ik_z^P z} - R_{PP}\frac{k_z^P}{k_P}e^{-ik_z^P z} - R_{PS}\frac{k_\rho}{k_S}e^{-ik_z^S z}. \label{eq:u_z_p}
\end{align}

An upward-incident SV wave produces reflected SV and P waves at the free surface \cite{AkiRichards2002}. The generalized parameters map to 
$s = \xi_0^{SV}$, $D_h(z) = U_\rho^{(SV)}(z)$, and $D_z(z) = U_z^{(SV)}(z)$, yielding:
\begin{align}
    U_\rho^{(SV)}(z) &= \frac{k_z^S}{k_S}e^{ik_z^S z} - R_{SS}\frac{k_z^S}{k_S}e^{-ik_z^S z} + R_{SP}\frac{k_\rho}{k_P}e^{-ik_z^P z}, \label{eq:u_rho_sv}\\
    U_z^{(SV)}(z) &= -\frac{k_\rho}{k_S}e^{ik_z^S z} - R_{SS}\frac{k_\rho}{k_S}e^{-ik_z^S z} - R_{SP}\frac{k_z^P}{k_P}e^{-ik_z^P z}. \label{eq:u_z_sv}
\end{align}
The reflection coefficients ($R_{PP}, R_{PS}, R_{SS}, R_{SP}$) follow the standard traction-free boundary conditions in $z=0$.

\begin{align}
R_{PP}
=
\frac{
4k_\rho^2 k_z^P k_z^S-(k_S^2-2k_\rho^2)^2
}{
(k_S^2-2k_\rho^2)^2+4k_\rho^2 k_z^P k_z^S
},
\qquad
R_{PS}
=
-\frac{
4(k_S/k_P)\,k_\rho k_z^P (k_S^2-2k_\rho^2)
}{
(k_S^2-2k_\rho^2)^2+4k_\rho^2 k_z^P k_z^S
},
\nonumber
\\
R_{SS}
=
\frac{
4k_\rho^2k_z^Pk_z^S-(k_S^2-2k_\rho^2)^2
}{
(k_S^2-2k_\rho^2)^2+4k_\rho^2k_z^Pk_z^S
},
\qquad
R_{SP}
=
\frac{
4(k_P/k_S)\,k_\rho k_z^S (k_S^2-2k_\rho^2)
}{
(k_S^2-2k_\rho^2)^2+4k_\rho^2k_z^Pk_z^S
}.
\end{align}
\subsubsection{Potential and Acceleration Mapping}

By linear superposition, depth-dependent transfer functions $\mathcal{B}_{\mathrm{pot}}(z_0)$ and $\mathcal{B}_{\mathrm{vert}}(z_0)$ are evaluated as sums of all contributing wave components $m$:
\begin{equation}
    \mathcal{B}_{\mathrm{pot}}(z_0) = \sum_m C_m R_m \, \mathcal{Z}_m(z_0), \qquad \mathcal{B}_{\mathrm{vert}}(z_0) = \sum_m C_m R_m \, \mathcal{V}_m(z_0).
\end{equation}

The intrinsic amplitude factors are $C_P = i/k_P$ and $C_S = 1/k_S$. The scattering factors are $R_m = 1$ for incident waves and $R_m \in \{R_{PP}, R_{PS}, R_{SS}, R_{SP}\}$ for reflected/converted components. 

The depth profiles $\mathcal{Z}_m(z_0)$ and $\mathcal{V}_m(z_0)$ are defined in a piecewise way:

\paragraph{(i) Above-ground test mass ($z_0 > 0$)}
\begin{align}
\mathcal{Z}(z_0) = e^{-k_\rho z_0},
\qquad
\mathcal{V}(z_0) = k_\rho e^{-k_\rho z_0}.
\end{align}

\paragraph{(ii) Underground test mass ($z_0 \le 0$)}
\begin{itemize}
    \item SV components:
    \begin{align}
        \mathcal{Z}_S(z_0) = e^{k_\rho z_0}, \qquad
        \mathcal{V}_S(z_0) = -k_\rho e^{k_\rho z_0}.
    \end{align}
    \item Upward-propagating P components:
    \begin{align}
        \mathcal{Z}_P^+(z_0) = 2 e^{i k_z^P z_0} - e^{k_\rho z_0}, \qquad
        \mathcal{V}_P^+(z_0) = -2 i k_z^P e^{i k_z^P z_0} + k_\rho e^{k_\rho z_0}.
    \end{align}
    \item Downward-propagating P components:
    \begin{align}
        \mathcal{Z}_P^-(z_0) = 2 e^{-i k_z^P z_0} - e^{k_\rho z_0}, \qquad
        \mathcal{V}_P^-(z_0) = 2 i k_z^P e^{-i k_z^P z_0} + k_\rho e^{k_\rho z_0}.
    \end{align}
\end{itemize}

\subsection{Rayleigh waves}

Rayleigh waves strictly propagate horizontally with the wavenumber $\mathbf{k}_h = \mathbf{k}_R = k_R(\cos\theta,\sin\theta,0)$. Their evanescent vertical decay is governed by $q_{zP} = \sqrt{k_R^2-k_P^2}$ and $q_{zS} = \sqrt{k_R^2-k_S^2}$, with the amplitude ratio $\zeta = \sqrt{q_{zP}/q_{zS}}$. \\
Due to their elliptical polarization, Rayleigh waves introduce a $90^\circ$ phase shift in the horizontal plane. We map the generalized parameters as $s = A$, $D_h(z) = i H(z)$, and $D_z(z) = V(z)$, where the Rayleigh depth profiles are \cite{Harms2019}:
\begin{align}
    H(z) = k_R e^{q_{zP}z} - \zeta q_{zS} e^{q_{zS}z}, \qquad
    V(z) = q_{zP} e^{q_{zP}z} - \zeta k_R e^{q_{zS}z}.
    \label{eq:R-depth}
\end{align}

Substituting $D_h(z) = i H(z)$ into the Master Correlation Tensor (Eq.~\eqref{eq:C_general}) analytically resolves the complex components (e.g., $i D'_h (D_z)^* = -H(z')V^*(z)$), naturally collapsing it into the purely real-valued Rayleigh tensor. \\
The gravitational transfer functions for Rayleigh waves are determined strictly by substituting the known exact potential fields into the Master Equations. The potential and acceleration coefficients can be mapped as follows:
\paragraph{(i) Above-ground test mass ($z_0 > 0$)}
Above the surface, the potential is governed by the evanescent decay of the Laplace domain $e^{-k_R z_0}$. Factoring this out dictates the profiles:
\begin{align}
    \mathcal{B}_{\mathrm{pot}}(z_0) = -(1-\zeta)e^{-k_R z_0}, \qquad
    \mathcal{B}_{\mathrm{vert}}(z_0) = -k_R(1-\zeta)e^{-k_R z_0}.
\end{align}

\paragraph{(ii) Underground test mass ($z_0 \le 0$)}
Underground, the potential is a superposition of the two distinct vertical decay modes ($q_{zP}$ and $k_R$). The mapped profiles are:
\begin{align}
    \mathcal{B}_{\mathrm{pot}}(z_0) = -2e^{q_{zP}z_0} + (1+\zeta)e^{k_R z_0}, \qquad
    \mathcal{B}_{\mathrm{vert}}(z_0) = 2q_{zP}e^{q_{zP}z_0} - (1+\zeta)k_R e^{k_R z_0}.
\end{align}

\section{Diffuse-field P- and S-wave fractions}
\label{app:modal_fractions}

For a three-dimensional diffuse elastic wavefield, modal equipartition gives
\begin{equation}
\frac{E_S}{E_P}
=2\left(\frac{\alpha}{\beta}\right)^3,
\label{eq:diffuse_energy_ratio}
\end{equation}
where the factor two accounts for the two independent shear polarizations
\cite{Weaver1982,Margerin2000}. To relate the restricted half-space P--SV
ensemble to the full-space P--S ensemble, define the common
one-shear-polarization P fraction
\begin{equation}
p\equiv\frac{E_P}{E_P+E_{S,1}},
\label{eq:common_p_definition}
\end{equation}
where \(E_{S,1}\) is the energy carried by one shear polarization. The
modal fields are compared at the same angular frequency and use a common
displacement normalization. Under this convention, their time-averaged
elastic energies are proportional to \(\rho_0\omega^2\) times their
displacement powers, so the energy fractions provide the spectral weights
used in the correlation tensors. The
half-space incident ensemble considered here contains P and SV waves but no SH
population. Its P fraction is therefore
\begin{equation}
p_{\mathrm{HS}}\equiv w_P=p,
\qquad
w_{SV}=1-p_{\mathrm{HS}}.
\label{eq:p_halfspace}
\end{equation}
Here \(p_{\mathrm{HS}}\) is the incident P-wave fraction within the restricted
P--SV ensemble. It is not the local P-wave fraction of the complete
free-surface field after reflection and P--SV mode conversion.
In full space, the S-wave population contains two statistically equivalent
transverse polarizations. The corresponding P fraction is
\begin{equation}
p_{\mathrm{FS}}
=\frac{E_P}{E_P+E_{SV}+E_{SH}}
=\frac{p_{\mathrm{HS}}}{2-p_{\mathrm{HS}}},
\qquad
p_{\mathrm{HS}}=\frac{2p_{\mathrm{FS}}}{1+p_{\mathrm{FS}}}.
\label{eq:p_hs_fs_relation}
\end{equation}
Thus, the two models implement the same equipartition-based modal-density prescription but have
different P fractions because they contain different numbers of shear
polarizations. Equivalently,
\begin{equation}
p_{\mathrm{HS}}
=\left[1+\left(\frac{\alpha}{\beta}\right)^3\right]^{-1},
\qquad
p_{\mathrm{FS}}
=\left[1+2\left(\frac{\alpha}{\beta}\right)^3\right]^{-1}.
\label{eq:equipartition_modal_fractions}
\end{equation}

\section{Body waves in full space}
\subsection{Analytical Body-Wave Models}
\label{ssec:bw_models}

Define the three-dimensional
separation vector and its magnitude by
\begin{equation}
  \mathbf{R} = \mathbf{r} - \mathbf{r}',
  \qquad
  R = |\mathbf{R}|,
  \qquad
  \hat{\mathbf{R}} = \frac{\mathbf{R}}{R},
  \label{eq:bw_sep}
\end{equation}
Although \(\hat{\mathbf R}\) is undefined at \(R=0\), all kernels below are
defined there by their continuous isotropic limit.

Following \cite{Harms2019}, introduce the P- and S-wave correlation kernels
normalized by their respective one-axis modal displacement spectra:
\begin{align}
\mathcal{K}_{ij}^{\mathrm{P}}(\mathbf{r}',\mathbf{r};\omega)
&=
\left[j_0(k_{\mathrm{P}}R)+j_2(k_{\mathrm{P}}R)\right]\delta_{ij}
-3j_2(k_{\mathrm{P}}R)\hat{R}_i\hat{R}_j,
\nonumber\\
\mathcal{K}_{ij}^{\mathrm{S}}(\mathbf{r}',\mathbf{r};\omega)
&=
\left[j_0(k_{\mathrm{S}}R)-\frac{1}{2}j_2(k_{\mathrm{S}}R)\right]\delta_{ij}
+\frac{3}{2}j_2(k_{\mathrm{S}}R)\hat{R}_i\hat{R}_j.
\label{eq:bw_f_tensor}
\end{align}
Both kernels satisfy
\(\mathcal{K}_{ij}^{A}(\mathbf{r},\mathbf{r};\omega)=\delta_{ij}\),
where \(A\in\{\mathrm{P},\mathrm{S}\}\).
Let \(S_{\mathrm{P}}(\omega)\equiv S(\xi_n^{\mathrm{P}};\omega)\) and
\(S_{\mathrm{S}}(\omega)\equiv S(\xi_n^{\mathrm{S}};\omega)\) denote the
one-axis modal spectra, and define
\(S_n=S_{\mathrm{P}}+S_{\mathrm{S}}\) and
\(p_{\mathrm{FS}}=S_{\mathrm{P}}/S_n\).
The modal correlations are
\(C_{ij}^{\mathrm{P}}=S_{\mathrm{P}}\mathcal{K}_{ij}^{\mathrm{P}}\)
and
\(C_{ij}^{\mathrm{S}}=S_{\mathrm{S}}\mathcal{K}_{ij}^{\mathrm{S}}\).
The mixed body-wave correlation tensor is therefore
\begin{equation}
C_{ij}(\mathbf{r}',\mathbf{r};\omega)
=S_n(\omega)\left[
p_{\mathrm{FS}}\,\mathcal{K}_{ij}^{\mathrm{P}}(\mathbf{r}',\mathbf{r};\omega)
+(1-p_{\mathrm{FS}})\,\mathcal{K}_{ij}^{\mathrm{S}}(\mathbf{r}',\mathbf{r};\omega)
\right].
\label{eq:bw_mix_tensor}
\end{equation}
When both modal spectra are nonzero, this is equivalently the ratio form
\(C_{ij}=S_n[(p_{\mathrm{FS}}/S_{\mathrm{P}})C_{ij}^{\mathrm{P}}
+((1-p_{\mathrm{FS}})/S_{\mathrm{S}})C_{ij}^{\mathrm{S}}]
=C_{ij}^{\mathrm{P}}+C_{ij}^{\mathrm{S}}\).
At zero separation, \(C_{ij}=S_n\delta_{ij}\).
If \(k_{\mathrm{P}}=k_{\mathrm{S}}=k\) and all three polarizations carry
equal power, then \(p_{\mathrm{FS}}=1/3\) and
\(C_{ij}=S_n j_0(kR)\delta_{ij}\), which provides a useful consistency
check.

\subsubsection{Exact Homogeneous Full-Space Model}
The corresponding closed-form gravitoelastic tensors for P- and S-waves are \cite{Harms2019}
\begin{equation}
c_{ij}^{(\mathrm{full},\mathrm{P})}(\mathbf{r},\omega)
=4\pi G\rho_0\,C_{ij}^{\mathrm{P}}(\mathbf{r}_0,\mathbf{r};\omega),
\qquad
c_{ij}^{(\mathrm{full},\mathrm{S})}(\mathbf{r},\omega)=0.
\label{eq:bw_full}
\end{equation}
S waves therefore produce no Newtonian acceleration in the full-space
geometry, a consequence of their purely transverse polarisation. Since
\(C_{ij}^{\mathrm{P}}\) already contains
  \(S_{\mathrm{P}}=p_{\mathrm{FS}}S_n\), the mixed result is
\begin{equation}
  c_{ij}^{(\mathrm{full})}(\mathbf{r},\omega)
    =4\pi G\rho_0\,
      C_{ij}^{\mathrm{P}}(\mathbf{r}_0,\mathbf{r};\omega).
  \label{eq:bw_mix_full}
\end{equation}

\subsubsection{Principal-Value Model}

Direct evaluation of the solid-angle integrals of Eq.~\eqref{eq:PV_def} yields

\begin{equation}
c_{ij}^{(\mathrm{PV},\mathrm{P})}(\mathbf{r},\omega)
=\frac{8\pi}{3}\,G\rho_0\,
C_{ij}^{\mathrm{P}}(\mathbf{r}_0,\mathbf{r};\omega),
\label{eq:bw_P_PV}
\end{equation}

\begin{equation}
c_{ij}^{(\mathrm{PV},\mathrm{S})}(\mathbf{r},\omega)
=-\frac{4\pi}{3}\,G\rho_0\,
C_{ij}^{\mathrm{S}}(\mathbf{r}_0,\mathbf{r};\omega),
\label{eq:bw_S_PV}
\end{equation}

and the mixed result is
\begin{equation}
  c_{ij}^{(\mathrm{PV})}(\mathbf{r},\omega)
    =G\rho_0\left[
      \frac{8\pi}{3}\,
      C_{ij}^{\mathrm{P}}(\mathbf{r}_0,\mathbf{r};\omega)
      -\frac{4\pi}{3}\,
      C_{ij}^{\mathrm{S}}(\mathbf{r}_0,\mathbf{r};\omega)
      \right].
  \label{eq:bw_mix_PV}
\end{equation}

\subsubsection{Spherical-Cavity Model}
Define
\begin{equation}
  M_{\mathrm{P}}^{(\mathrm{cav})}(a)
    = 8\pi\,\frac{j_1(k_{\mathrm{P}}a)}{k_{\mathrm{P}}a},
  \qquad
  M_{\mathrm{S}}^{(\mathrm{cav})}(a)
    = -4\pi\,\frac{j_1(k_{\mathrm{S}}a)}{k_{\mathrm{S}}a},
  \label{eq:bw_coeffs_cav}
\end{equation}
where $j_1$ is the spherical Bessel function of order~1, consistent with the
spherical functions $j_0$ and $j_2$ in the full-space correlation kernels.
These should not be confused with the cylindrical Bessel functions $J_n$
appearing in the horizontally isotropic half-space tensors. Therefore,
\begin{equation}
c_{ij}^{(\mathrm{cav},\mathrm{P})}(\mathbf{r},\omega)
= G\rho_0\,M_{\mathrm{P}}^{(\mathrm{cav})}(a)\,
C_{ij}^{\mathrm{P}}(\mathbf{r}_0,\mathbf{r};\omega),
\label{eq:bw_P_cav}
\end{equation}

\begin{equation}
c_{ij}^{(\mathrm{cav},\mathrm{S})}(\mathbf{r},\omega)
= G\rho_0\,M_{\mathrm{S}}^{(\mathrm{cav})}(a)\,
C_{ij}^{\mathrm{S}}(\mathbf{r}_0,\mathbf{r};\omega),
\label{eq:bw_S_cav}
\end{equation}

and the mixed result is
\begin{equation}
  c_{ij}^{(\mathrm{cav})}(\mathbf{r},\omega)
    =G\rho_0\left[
      M_{\mathrm{P}}^{(\mathrm{cav})}(a)\,
      C_{ij}^{\mathrm{P}}(\mathbf{r}_0,\mathbf{r};\omega)
      +M_{\mathrm{S}}^{(\mathrm{cav})}(a)\,
      C_{ij}^{\mathrm{S}}(\mathbf{r}_0,\mathbf{r};\omega)
      \right].
  \label{eq:bw_mix_cav}
\end{equation}

\section{Angular Bessel identities}\label{app:bessel-identities}

In the angular integrations below, \(\theta\) denotes the propagation angle of a generic horizontal wavevector
\(
\mathbf{k}_h(\theta)
=
\kappa(\cos\theta,\sin\theta),
\)
where \(\kappa\) is the corresponding horizontal wavenumber. The angle \(\varphi\) denotes the polar angle of a generic horizontal separation vector 
\(
\Delta\bm{\rho}
=
\bm{\rho}'-\bm{\rho}
=
\Delta\rho(\cos\varphi,\sin\varphi).
\)
We define
\(
q=\kappa\Delta\rho,
\)
so that
\(
\mathbf{k}_h(\theta)\cdot\Delta\bm{\rho}
=
q\cos(\theta-\varphi).
\)
This notation is independent of the wave type. For example, in the Rayleigh-wave case one sets \(\kappa=k_R\), while in a body-wave calculation \(\kappa\) may represent the relevant horizontal wavenumber.

Repeated use is made of the angular identities
\begin{equation}
\begin{aligned}
\frac{1}{2\pi}\int_0^{2\pi} e^{iq\cos(\theta-\varphi)}\,d\theta
&= J_0(q),\\
\frac{1}{2\pi}\int_0^{2\pi} \cos\theta\,e^{iq\cos(\theta-\varphi)}\,d\theta
&= i\cos\varphi\,J_1(q),\\
\frac{1}{2\pi}\int_0^{2\pi} \sin\theta\,e^{iq\cos(\theta-\varphi)}\,d\theta
&= i\sin\varphi\,J_1(q),\\
\frac{1}{2\pi}\int_0^{2\pi} \cos^2\theta\,e^{iq\cos(\theta-\varphi)}\,d\theta
&= \frac12\!\left[J_0(q)-\cos(2\varphi)\,J_2(q)\right],\\
\frac{1}{2\pi}\int_0^{2\pi} \sin^2\theta\,e^{iq\cos(\theta-\varphi)}\,d\theta
&= \frac12\!\left[J_0(q)+\cos(2\varphi)\,J_2(q)\right],\\
\frac{1}{2\pi}\int_0^{2\pi} \sin\theta\cos\theta\,e^{iq\cos(\theta-\varphi)}\,d\theta
&= -\frac12\sin(2\varphi)\,J_2(q).
\end{aligned}
\label{eq:bessel_angular_identities}
\end{equation}

For the reversed horizontal phase, \(q\to -q\), the required parity relations are
\begin{equation}
J_0(-q)=J_0(q),
\qquad
J_1(-q)=-J_1(q),
\qquad
J_2(-q)=J_2(q).
\label{eq:bessel_parity}
\end{equation}

\end{appendices}

\bibliographystyle{iopart-num}
\bibliography{sn-bibliography,nn_references}

\end{document}